\documentclass[11pt,twoside]{article}

\usepackage{amsfonts,epsfig,graphicx,subfigure}
\usepackage{amsmath,amssymb,amsthm}
\usepackage{fullpage}
\usepackage[normalem]{ulem}
\usepackage{epic,eepic}

\usepackage{epsf}
\usepackage{graphics}
\usepackage{epic}
\usepackage{eepic}
\usepackage{eepicemu}
\usepackage{psfrag}
\usepackage{hyperref}

\newcommand{\opnorm}[1]{\ensuremath{\matsnorm{#1}{\myop}}}
\newcommand{\matsnormsq}[2]{|\!|\!| #1 | \! | \!|^2_{{#2}}}
\newcommand{\myop}{\ensuremath{\operatorname{op}}}

\theoremstyle{plain}

\newtheorem{theo}{Theorem}[section]

\newtheorem{lem}{Lemma}[section]
\newtheorem{prop}{Proposition}[section]
\newtheorem{cor}{Corollary}[section]

\theoremstyle{definition} 

\newtheorem{nota}{Notation}[section]
\newtheorem{de}{Definition}[section]
\newtheorem{exa}{Example}[section]
\newtheorem{as}{Assumption}[section]
\newtheorem{alg}{Algorithm}[section]

\newcommand{\btheo}{\begin{theo}}
\newcommand{\bde}{\begin{de}}
\newcommand{\ble}{\begin{lem}}
\newcommand{\bpr}{\begin{prop}}
\newcommand{\bno}{\begin{nota}}
\newcommand{\bex}{\begin{exa}}
\newcommand{\bcor}{\begin{cor}}
\newcommand{\spro}{\begin{proof}}
\newcommand{\bas}{\begin{as}}
\newcommand{\balg}{\begin{alg}}

\newcommand{\etheo}{\end{theo}}
\newcommand{\ede}{\end{de}}
\newcommand{\ele}{\end{lem}}
\newcommand{\epr}{\end{prop}}
\newcommand{\eno}{\end{nota}}
\newcommand{\eex}{\end{exa}}
\newcommand{\ecor}{\end{cor}}
\newcommand{\fpro}{\end{proof}}
\newcommand{\eas}{\end{as}}
\newcommand{\ealg}{\end{alg}}

\theoremstyle{plain}

\newtheorem{theos}{Theorem}
\newtheorem{props}{Proposition}
\newtheorem{lems}{Lemma}
\newtheorem{cors}{Corollary}

\theoremstyle{definition}
\newtheorem{exas}{Example}
\newtheorem{algs}{Algorithm}
\newtheorem{asss}{Assumption}
\newtheorem{defns}{Definition}

\newcommand{\btheos}{\begin{theos}}
\newcommand{\etheos}{\end{theos}}
\newcommand{\bprops}{\begin{props}}
\newcommand{\eprops}{\end{props}}
\newcommand{\bdes}{\begin{defns}}
\newcommand{\edes}{\end{defns}}
\newcommand{\blems}{\begin{lems}}
\newcommand{\elems}{\end{lems}}
\newcommand{\bcors}{\begin{cors}}
\newcommand{\ecors}{\end{cors}}
\newcommand{\bexs}{\begin{exas}}
\newcommand{\eexs}{\end{exas}}
\newcommand{\balgs}{\begin{algs}}
\newcommand{\ealgs}{\end{algs}}
\newcommand{\bass}{\begin{asss}}
\newcommand{\eass}{\end{asss}}

\newlength{\widebarargwidth}
\newlength{\widebarargheight}
\newlength{\widebarargdepth}
\DeclareRobustCommand{\widebar}[1]{%
  \settowidth{\widebarargwidth}{\ensuremath{#1}}%
  \settoheight{\widebarargheight}{\ensuremath{#1}}%
  \settodepth{\widebarargdepth}{\ensuremath{#1}}%
  \addtolength{\widebarargwidth}{-0.3\widebarargheight}%
  \addtolength{\widebarargwidth}{-0.3\widebarargdepth}%
  \makebox[0pt][l]{\hspace{0.3\widebarargheight}%
    \hspace{0.3\widebarargdepth}%
    \addtolength{\widebarargheight}{0.3ex}%
    \rule[\widebarargheight]{0.95\widebarargwidth}{0.1ex}}%
  {#1}}

\newcommand{\matsnorm}[2]{|\!|\!| #1 | \! | \!|_{{#2}}}

\newenvironment{carlist}
 {\begin{list}{$\bullet$}
 {\setlength{\topsep}{0in} \setlength{\partopsep}{0in}
  \setlength{\parsep}{0in} \setlength{\itemsep}{\parskip}
  \setlength{\leftmargin}{0.07in} \setlength{\rightmargin}{0.08in}
  \setlength{\listparindent}{0in} \setlength{\labelwidth}{0.08in}
  \setlength{\labelsep}{0.1in} \setlength{\itemindent}{0in}}}
 {\end{list}}

\newcommand{\bcar}{\begin{carlist}}
\newcommand{\ecar}{\end{carlist}}

\newcommand{\tracer}[2]{\ensuremath{\langle \!\langle {#1}, \; {#2}
\rangle \!\rangle}}

\newcommand{\vmax}{\ensuremath{\vee}}

\makeatletter
\long\def\@makecaption#1#2{
        \vskip 0.8ex
        \setbox\@tempboxa\hbox{\small {\bf #1:} #2}
        \parindent 1.5em  
        \dimen0=\hsize
        \advance\dimen0 by -3em
        \ifdim \wd\@tempboxa >\dimen0
                \hbox to \hsize{
                        \parindent 0em
                        \hfil 
                        \parbox{\dimen0}{\def\baselinestretch{0.96}\small
                                {\bf #1.} #2
                                } 
                        \hfil}
        \else \hbox to \hsize{\hfil \box\@tempboxa \hfil}
        \fi
        }
\makeatother

\long\def\comment#1{}

\makeatletter
\def\@cite#1#2{[\if@tempswa #2 \fi #1]}
\makeatother

\makeatletter
\long\def\barenote#1{
    \insert\footins{\footnotesize
    \interlinepenalty\interfootnotelinepenalty 
    \splittopskip\footnotesep
    \splitmaxdepth \dp\strutbox \floatingpenalty \@MM
    \hsize\columnwidth \@parboxrestore
    {\rule{\z@}{\footnotesep}\ignorespaces
      #1\strut}}}
\makeatother

\newcommand{\bit}{\begin{itemize}}
\newcommand{\eit}{\end{itemize}}
\newcommand{\ben}{\begin{enumerate}}
\newcommand{\een}{\end{enumerate}}

\newcommand{\bear}{\begin{eqnarray}}
\newcommand{\eear}{\end{eqnarray}}

\newcommand{\order}{{\mathcal{O}}}

\newcommand{\inprod}[2]{\ensuremath{\langle #1 , \, #2 \rangle}}

\newcommand{\kull}[2]{\ensuremath{D(#1\; \| \; #2)}}

\newcommand{\ones}{\ensuremath{\mathbf{1}}}

\newcommand{\Exs}{\ensuremath{{\mathbb{E}}}}

\newcommand{\beq}{\begin{quotation}}
\newcommand{\enq}{\end{quotation}}

\newcommand{\estart}{\begin{equation}}
\newcommand{\eend}{\end{equation}}

\newcommand{\widgraph}[2]{\includegraphics[keepaspectratio,width=#1]{#2}}

\newcommand{\defn}{\ensuremath{:  =}}

\newcommand{\ysca}{{{y}}}

\newcommand{\bec}{\begin{center}}
\newcommand{\enc}{\end{center}}

\newcommand{\beit}{\begin{itemize}}
\newcommand{\enit}{\end{itemize}}

\newcommand{\been}{\begin{enumerate}}
\newcommand{\enen}{\end{enumerate}}

\newcommand{\comsl}{\begin{slide}}
\newcommand{\comspor}{\begin{slide*}}

\newcommand{\comsld}[2]{\begin{slide}[#1,#2]}
\newcommand{\comspord}[2]{\begin{slide*}[#1,#2]}

\newcommand{\mendsl}{\end{slide}}
\newcommand{\mendspo}{\end{slide*}}

\newcommand{\wtil}[1]{\ensuremath{\widetilde{#1}}}

\newcommand{\real}{\ensuremath{{\mathbb{R}}}}

\newcommand{\numobs}{\ensuremath{n}}
\newcommand{\pdim}{\ensuremath{p}}

\newcommand{\regpar}{\ensuremath{\lambda_\numobs}}

\newcommand{\Ball}{\ensuremath{\mathbb{B}}}

\newcommand{\Regone}[1]{\ensuremath{\|#1\|_1}}
\newcommand\Errone\Regone

\newcommand{\qpar}{\ensuremath{q}}

\newcommand{\rdim}{\ensuremath{r}}
\newcommand{\mdim}{\ensuremath{d}}

\newcommand{\mprob}{\ensuremath{\mathbb{P}}}

\newcommand{\radplain}{\ensuremath{\rho}}
\newcommand{\radq}{\ensuremath{\radplain_\qpar}}

\def\real{\mathbb{R}}

\def\defn{:=}

\newcommand{\frobnorm}[1]{\ensuremath{\matsnorm{#1}{F}}}

\newcommand{\Xop}{\ensuremath{{\mathfrak{X}_\numobs}}}

\newcommand{\lra}{\ensuremath{\mdim_1}}
\newcommand{\lrb}{\ensuremath{\mdim_2}}
\newcommand{\mdima}{\lra}
\newcommand{\mdimb}{\lrb}

\newcommand{\plaincon}{\ensuremath{c}}

\newcommand{\myrad}{\ensuremath{D}}

\newcommand{\ObsMat}[1]{\ensuremath{X^{(#1)}}}
\newcommand{\ObsTil}[1]{\ensuremath{\widetilde{X}^{(#1)}}}
\newcommand{\ObsY}[1]{{\ensuremath{Y^{(#1)}}}}

\newcommand{\SuperSum}[1]{\ensuremath{\frac{\|\XopNew(#1)\|_2}{\sqrt{\numobs}}}}

\newcommand{\SuperSumSq}[1]{\ensuremath{\frac{1}{\numobs}
\sum_{i=1}^n \tracer{\ObsTil{i}}{#1}^2}}

\newcommand{\newrade}[1]{\ensuremath{\varepsilon_{#1}}}
\newcommand{\rade}{\ensuremath{\varepsilon}}

\newcommand{\LowerMess}{\ensuremath{\sqrt{\frac{\MSUM \log
        \MSUM}{\numobs}}}}

\newcommand{\SpecRadPlain}{\ensuremath{\rho}}
\newcommand{\SpecRad}{\ensuremath{\SpecRadPlain(\myrad)}}

\newcommand{\alspike}{\ensuremath{\alpha_{\operatorname{sp}}}}
\newcommand{\alrank}{\ensuremath{\beta_{\operatorname{ra}}}}

\newcommand{\EvilEvent}{\ensuremath{\mathcal{E}}}

\newcommand{\ThetaInt}{\ensuremath{\widetilde{\Theta}}}
\newcommand{\Event}{\ensuremath{\mathcal{E}}}

\newcommand{\ThetaHat}{\ensuremath{\widehat{\Theta}}}
\newcommand{\ThetaStar}{\ensuremath{\Theta^*}}

\newcommand{\japan}{\ensuremath{c_5}}

\newcommand{\Ballspec}{\ensuremath{\widetilde{\Ball}}}
\newcommand{\PackNum}{\ensuremath{M}}
\newcommand{\Lind}{\ensuremath{V}}
\newcommand{\Lhat}{\ensuremath{\widehat{V}}}
\newcommand{\Coke}{\ensuremath{\mathbb{S}}}

\newcommand{\mrow}{{\ensuremath{\mdim_r}}}
\newcommand{\mcol}{\ensuremath{{\mdim_c}}}

\newcommand{\prow}{{\ensuremath{\mdim}}}
\newcommand{\pcol}{\ensuremath{\mdim}}

\newcommand{\Row}{\ensuremath{R}}
\newcommand{\Col}{\ensuremath{C}}

\newcommand{\Weifro}[1]{\ensuremath{\matsnorm{#1}{{\omega(F)}}}}

\newcommand{\Weifrosq}[1]{\ensuremath{|\!|\!| #1 | \! | \!|^2_{\omega(F)}}}

\newcommand{\Weinuc}[1]{\ensuremath{\matsnorm{#1}{{\omega(1)}}}}
\newcommand{\Weiplinf}[1]{\ensuremath{\matsnorm{#1}{{\omega(\infty)}}}}

\newcommand{\GamPlain}{\ensuremath{\Gamma}}
\newcommand{\altwospike}{\ensuremath{\alspike'}}
\newcommand{\altworank}{\ensuremath{\alrank'}}

\newcommand{\SpecSet}{\ensuremath{\mathfrak{B}}}
\newcommand{\SpecSetTil}{\ensuremath{\widebar{\SpecSet}}}

\newcommand{\AnnoySet}{\ensuremath{\mathfrak{D}(\delta, R)}}
\newcommand{\AnnoySetTwo}[1]{\ensuremath{\mathfrak{D}(#1, R)}}
\newcommand{\AnnoySup}[1]{\ensuremath{\sup_{#1\in \AnnoySet}}}
\newcommand{\AnnoySupTwo}[2]{\ensuremath{\sup_{#1\in \AnnoySetTwo{#2}}}}

\newcommand{\XopNew}{\ensuremath{\Xop'}}

\newcommand{\TEA}{\ensuremath{48}}

\newcommand{\LeadConHalf}{\ensuremath{\TEA L}}

\newcommand{\Lcon}{\ensuremath{L}}

\newcommand{\Ntil}{\ensuremath{\widetilde{N}}}

\newcommand{\TempSet}{\ensuremath{\mathcal{S}}}

\newcommand{\MTIMES}{\ensuremath{\sqrt{\mrow \mcol}}}

\newcommand{\MTIMESPOST}{\ensuremath{\mdim}}
\newcommand{\MSUMPOST}{\ensuremath{\mdim}}

\newcommand{\MSUM}{\ensuremath{\mdim}}

\newcommand{\Constraint}{\ensuremath{\mathfrak{C}}}
\newcommand{\ConstraintNew}{\ensuremath{\mathfrak{C}'}}

\newcommand{\THMCON}{\ensuremath{\Constraint(\numobs; \Kcon)}}

\newcommand{\THMCONNEW}{\ensuremath{\ConstraintNew(\numobs; \Kcon)}}

\newcommand{\ytil}{\ensuremath{\widetilde{y}}}

\newcommand{\noisevar}{\ensuremath{\nu}}

\newcommand{\spikeupper}{\ensuremath{\alpha^*}}
\newcommand{\upperspike}{\ensuremath{\spikeupper}}

\newcommand{\DeltaHat}{\ensuremath{\widehat{\Delta}}}
\newcommand{\Kcon}{\ensuremath{c_0}}

\newcommand{\newvsca}{\ensuremath{\xi}}

\newcommand{\vscatil}{\ensuremath{\widetilde{\newvsca}}}

\newcommand{\rank}{\ensuremath{\operatorname{rank}}}

\newcommand{\PlError}{\ensuremath{\DeltaHat}}

\newcommand{\mata}{\ensuremath{U}} 
\newcommand{\matb}{\ensuremath{V}}

\newcommand{\PlErrorA}{\ensuremath{\PlError'}}
\newcommand{\PlErrorB}{\ensuremath{\PlError''}}

\newcommand{\mataspec}{\ensuremath{\widetilde{\mata}}}
\newcommand{\matbspec}{\ensuremath{\widetilde{\matb}}}

\newcommand{\nucnorm}[1]{\matsnorm{#1}{1}} 

\newcommand{\GammaStar}{\ensuremath{\Gamma^*}}
\newcommand{\GammaHat}{\ensuremath{\widehat{\Gamma}}}

\newcommand{\DeltaTil}{\ensuremath{\widetilde{\Delta}}}

\newcommand{\lamstar}{\ensuremath{\regpar^*}}

\newcommand{\TAILSUM}{\ensuremath{\sum_{j = \rdim+1}^\mrow
\sigma_j(\GammaStar)}}

\newcommand{\MiniMax}{\ensuremath{\mathfrak{M}}}

\newcommand{\ThetaTil}{\ensuremath{\widetilde{\Theta}}}

\newcommand{\sigmax}{\ensuremath{\sigma_{\operatorname{max}}}}
\newcommand{\sigmin}{\ensuremath{\sigma_{\operatorname{min}}}}

\begin{document}

\begin{center}

{\bf{\LARGE{Restricted strong convexity and weighted matrix
      completion: Optimal bounds with noise}}}

\vspace*{.2in}

{\large{
\begin{tabular}{ccc}
Sahand Negahban & &  Martin J. Wainwright$^{\dagger,\star}$ 
\end{tabular}
}}

\vspace*{.2in}

September 2010

\vspace*{.2in}

\begin{tabular}{c}
Technical Report, \\
Department of Statistics,  UC Berkeley
\end{tabular}

\vspace*{.1in}

\begin{abstract}
We consider the matrix completion problem under a form of row/column
weighted entrywise sampling, including the case of uniform entrywise
sampling as a special case.  We analyze the associated random
observation operator, and prove that with high probability, it
satisfies a form of restricted strong convexity with respect to
weighted Frobenius norm.  Using this property, we obtain as
corollaries a number of error bounds on matrix completion in the
weighted Frobenius norm under noisy sampling and for both exact and
near low-rank matrices.  Our results are based on measures of the
``spikiness'' and ``low-rankness'' of matrices that are less
restrictive than the incoherence conditions imposed in previous work.
Our technique involves an $M$-estimator that includes controls on both
the rank and spikiness of the solution, and we establish
non-asymptotic error bounds in weighted Frobenius norm for recovering
matrices lying with $\ell_\qpar$-``balls'' of bounded spikiness.
Using information-theoretic methods, we show that no algorithm can
achieve better estimates (up to a logarithmic factor) over these same
sets, showing that our conditions on matrices and associated rates are
essentially optimal.
\end{abstract}

\end{center}


\section{Introduction}

Matrix completion problems correspond to reconstructing matrices,
either exactly or approximately, based on observing a subset of their
entries~\cite{Lau01,Deza97}.  In the simplest formulation of matrix
completion, the observations are assumed to be uncorrupted, whereas a
more general formulation (as considered in this paper) allows for
noisiness in these observations.  Matrix recovery based on only
partial information is an ill-posed problem, and accurate estimates
are possible only if the matrix satisfies additional structural
constraints, with examples including bandedness, positive
semidefiniteness, Euclidean distance measurements, Toeplitz, and
low-rank structure (see the survey paper~\cite{Lau01} and references
therein for more background).

The focus of this paper is low-rank matrix completion based on noisy
observations.  This problem is motivated by a variety of applications
where an underlying matrix is likely to have low-rank, or near
low-rank structure.  The archetypal example is the Netflix challenge,
a version of the collaborative filtering problem, in which the unknown
matrix is indexed by individuals and movies, and each observed entry
of the matrix corresponds to the rating assigned to the associated
movie by the given individual.  Since the typical person only watches
a tiny number of movies (compared to the total Netflix database), it
is only a sparse subset of matrix entries that are observed.  In this
context, one goal of collaborative filtering is to use the observed
entries to make recommendations to a person regarding movies that they
have \emph{not} yet seen.  We refer the reader to Srebro's
thesis~\cite{Sr04} (and references therein) for further discussion and
motivation for collaborative filtering and related problems.

In this paper, we analyze a method for approximate low-rank matrix
recovery using an $M$-estimator that is a combination of a data term,
and a weighted nuclear norm as a regularizer.  The nuclear norm is the
sum of the singular values of a matrix~\cite{Horn85}, and has been
studied in a body of past work, both on matrix completion and more
general problems of low-rank matrix estimation
(e.g.,~\cite{Fa02,Sr04,SreAloJaa05,SreRenJaa04,RecFazPar07,Bach08b,
  CanTao09,Recht09,Keshavan09a,Keshavan09b,NegWai09,RohTsy10}).  A
parallel line of work has studied computationally efficient algorithms
for solving problems with nuclear norm constraints
(e.g,~\cite{MazHasTib09, Nesterov07,LinEtAl09b}).  Here we limit our
detailed discussion to those papers that study various aspects of the
matrix completion problem.  Motivated by various problems in
collaborative filtering, Srebro and
colleagues~\cite{Sr04,SreAloJaa05,SreRenJaa04} studied various aspects
nuclear norm regularization; among various other contributions, Srebro
et al.~\cite{SreAloJaa05} established generalization error bounds
under certain conditions.  Candes and Recht~\cite{CanRec08} studied
the exact reconstruction of a low-rank matrix given perfect
(noiseless) observations of a subset of entries, and provided
sufficient conditions for exact recovery via nuclear norm relaxation.
These results were then refined in follow-up
work~\cite{CanTao09,Recht09}, with the simplest approach to date being
provided by Recht~\cite{Recht09}.  In a parallel line of work,
Keshavan et al.~\cite{Keshavan09a,Keshavan09b} have studied a method
based on thresholding and singular value decomposition, and
established various results on its behavior, both for noiseless and
noisy matrix completion.  Among other results, Rohde and
Tsybakov~\cite{RohTsy10} establish prediction error bounds for matrix
completion, a different metric than the matrix recovery problem of
interest here.  In recent work, Salakhutdinov and
Srebro~\cite{SalSre10} provided various motivations for the use of
weighted nuclear norms, in particular showing that the standard
nuclear norm relaxation can behave very poorly when the sampling is
non-uniform.  The analysis of this paper applies to both uniform and
non-uniform sampling, as well as a form of reweighted nuclear norm as
suggested by these authors, one which includes the ordinary nuclear
norm as a special case.  We provide a more detailed comparison between
our results and some aspects of past work in Section~\ref{SecCompare}.

As has been noted before~\cite{CanPla09}, a significant theoretical
challenge is that conditions that have proven very useful for sparse
linear regression---among them the restricted isometry property---are
\emph{not} satisfied for the matrix completion problem.  For this
reason, it is natural to seek an alternative and less restrictive
property that might be satisfied in the matrix completion setting.  In
recent work, Negahban et al.~\cite{NegRavWaiYu09} have isolated a
weaker condition known as \emph{restricted strong convexity} (RSC),
and proven that certain statistical models satisfy RSC with high
probability when the associated regularizer satisfies a
\emph{decomposability} condition.  When an $M$-estimator satisfies the
RSC condition, it is relatively straightforward to derive
non-asymptotic error bounds on parameter
estimates~\cite{NegRavWaiYu09}. The class of decomposable regularizers
includes the nuclear norm as particular case, and the
RSC/decomposability approach has been exploited to derive bounds for
various matrix estimation problems, among them multi-task learning,
autoregressive system identification, and compressed
sensing~\cite{NegWai09}.

To date, however, an open question is whether or not an appropriate
form of RSC holds for the matrix completion problem.  If it did hold,
then it would be possible to derive non-asymptotic error bounds (in
Frobenius norm) for matrix completion based on noisy observations.
Within this context, the main contribution of this paper is to prove
that with high probability, a form of the RSC condition holds for the
matrix completion problem, in particular over an interesting set of
matrices $\Constraint$, as defined in
equation~\eqref{EqnDefnConstraint} to follow, that have both low
nuclear/Frobenius norm ratio and low ``spikiness''.  The set
$\Constraint$ also excludes a neighborhood around zero, which is
essential so as to eliminate the nullspace of the sampling operator
underlying matrix completion.  Exploiting this RSC condition then
allows us to derive non-asymptotic error bounds on matrix recovery in
weighted Frobenius norms, both for exactly and approximately low-rank
matrices.  The theoretical core of this paper consists of three main
results.  Our first result (Theorem~\ref{ThmRSC}) proves that the
matrix completion loss function satisfies restricted strong convexity
with high probability over the set $\Constraint$.  Our second result
(Theorem~\ref{ThmMatrixEst}) exploits this fact to derive a
non-asymptotic error bound for matrix recovery in the weighted
Frobenius norm, one applicable to general matrices.  We then
specialize this result to the problem of estimating exactly low-rank
matrices (with a small number of non-zero singular values), as well as
near low-rank matrices characterized by relatively swift decay of
their singular values.  To the best of our knowledge, our results on
near low-rank matrices are the first for approximate matrix recovery
in the noisy setting, and as we discuss at more length in
Section~\ref{SecCompare}, our results on the exactly low-rank case are
sharper than past work on the problem.  Indeed, our final result
(Theorem~\ref{ThmLower}) uses information-theoretic techniques to
establish that up to logarithmic factors, no algorithm can obtain
faster rates than our method over the $\ell_\qpar$-balls of matrices
with bounded spikiness treated in this paper.

The remainder of this paper is organized as follows.  We begin in
Section~\ref{SecBackground} with background and a precise formulation
of the problem.  Section~\ref{SecMain} is devoted to a statement of
our main results, and discussion of some of their consequences.  In
Sections~\ref{SecProofThmMatrixEst} and Section~\ref{SecProofThmRSC},
we prove our main results, with more technical aspects of the
arguments deferred to appendices.  We conclude with a discussion in
Section~\ref{SecDiscuss}.


\section{Background and problem formulation}
\label{SecBackground}

In this section, we introduce background on low-rank matrix completion
problem, and also provide a precise statement of the problem studied
in this paper.

\subsection{Uniform and weighted sampling models}

Let $\ThetaStar \in \real^{\mrow \times \mcol}$ be an unknown matrix,
and consider an observation model in which we make $\numobs$
i.i.d. observations of the form
\begin{align}
\label{EqnOrigObs}
\ytil_i & = \ThetaStar_{j(i) k(i)} + \frac{\noisevar}{\MTIMES}
\vscatil_i,
\end{align}
Here the quantities $\frac{\noisevar}{\MTIMES} \vscatil_i$ correspond
to additive observation noises with variance appropriately scaled
according to the matrix dimensions. In defining the observation model,
one can either allow the Frobenius norm of $\ThetaStar$ to grow with
the dimension, as in done in other work~\cite{CanPla09,Keshavan09b},
or rescale the noise as we have done here.  This choice is consistent
with our assumption that $\ThetaStar$ has constant Frobenius norm
regardless of its rank or dimensions.  With this scaling, each
observation in the model~\eqref{EqnOrigObs} has a constant
signal-to-noise ratio regardless of matrix dimensions.

 In the simplest model, the row $j(i)$ and column $k(i)$ indices are
chosen uniformly at random from the sets $\{1, 2, \ldots, \mrow\}$ and
$\{1, 2, \ldots, \mcol \}$ respectively.  In this paper, we consider a
somewhat more general weighted sampling model.  In particular, let
\mbox{$\Row \in \real^{\mrow \times \mrow}$} and \mbox{$\Col \in
\real^{\mcol \times \mcol}$} be diagonal matrices, with rescaled
diagonals $\{\Row_j/\mrow, j = 1, 2, \ldots, \mrow\}$ and
$\{\Col_k/\mcol, k = 1, 2, \ldots, \mcol \}$ representing probability
distributions over the rows and columns of an $\mrow \times \mcol$
matrix.  We consider the weighted sampling model in which we make a
noisy observation of entry $(j,k)$ with probability $\Row_j
\Col_k/(\mrow \mcol)$, meaning that the row index $j(i)$ (respectively
column index $k(i)$) is chosen according to the probability
distribution $\Row/\mrow$ (respectively $\Col/\mcol$).  Note that in
the special case that $\Row = \ones_\mrow$ and $\Col = \ones_\mcol$,
the observation model~\eqref{EqnOrigObs} reduces to the usual model of
uniform sampling.

We assume that each row and column is sampled with positive
probability, in particular that there is some constant $1 \leq \Lcon <
\infty$ such that $\Row_a \geq 1/\Lcon$ and $\Col_b \geq 1/\Lcon$ for
all rows and columns.  However, apart from the constraints
$\sum_{a=1}^\mrow \Row_{aa} = \mrow$ and \mbox{$\sum_{b=1}^\mcol
  \Col_{bb} = \mcol$,} we do not require that the row and column
weights remain bounded as $\mrow$ and $\mcol$ tend to infinity.

\subsection{The observation operator and restricted strong convexity}

We now describe an alternative formulation of the observation
model~\eqref{EqnOrigObs} that, while statistically equivalent to the
original, turns out to be more natural for analysis.  For each $i = 1,
2, \ldots, \numobs$, define the matrix
\begin{align}
\ObsMat{i} & = \MTIMES \, \newrade{i} \; e_{a(i)} e_{b(i)}^T,
\end{align}
where $\newrade{i} \in \{-1, +1\}$ is a random sign, and consider the
observation model
\begin{align}
\label{EqnNetflix}
\ysca_i & = \tracer{\ObsMat{i}}{\ThetaStar} + \noisevar \, \newvsca_i,
  \qquad \mbox{ for $i = 1, \ldots, \numobs$,}
\end{align}
where $\tracer{A}{B} \defn \sum_{j,k} A_{jk} B_{jk}$ is the trace
inner product, and $\newvsca_i$ is an additive noise from the same
distribution as the original model.  The model~\eqref{EqnNetflix} is
can be obtained from the original model~\eqref{EqnOrigObs} by
rescaling all terms by the factor $\MTIMES$, and introducing the
random signs $\newrade{i}$.  The rescaling has no statistical effect,
and nor do the random signs, since the noise is symmetric (so that
$\newvsca_i = \newrade{i} \vscatil_i$ has the same distribution as
$\vscatil_i$).  Thus, the observation model~\eqref{EqnNetflix} is
statistically equivalent to the original one~\eqref{EqnOrigObs}.

  In order to specify a vector form of the observation model, let us
define an operator \mbox{$\Xop: \real^{\mrow \times \mcol} \rightarrow
\real^\numobs$} via
\begin{align*}
[\Xop(\Theta)]_i & \defn \tracer{\ObsMat{i}}{\Theta}, \quad \mbox{for
  $i = 1, 2, \ldots \numobs$.}
\end{align*}
We refer to $\Xop$ as the \emph{observation operator}, since it maps
any matrix $\Theta \in \real^{\mrow \times \mcol}$ to an
$\numobs$-vector of samples.   With this notation, we can write
the observations~\eqref{EqnNetflix} in a vectorized form as
$y = \Xop(\ThetaStar) + \noisevar \newvsca$.

The reformulation~\eqref{EqnNetflix} is convenient for various
reasons.  For any matrix $\Theta \in \real^{\mrow \times \mcol}$, we
have $\Exs [\tracer{\ObsMat{i}}{\Theta}] = 0$ and
\begin{align}
\Exs \big[ \tracer{\ObsMat{i}}{\Theta}^2 \big] & = \sum_{j=1}^\mrow
\sum_{k=1}^\mcol \Row_{j} \Theta_{jk}^2 \Col_k \; = \;
\underbrace{\matsnormsq{\sqrt{\Row} \Theta
\sqrt{\Col}}{F}}_{{\Weifrosq{\Theta}}},
\end{align}
where we have defined the \emph{weighted Frobenius norm}
$\Weifro{\cdot}$ in terms of the row $\Row$ and column $\Col$ weights.
As a consequence, the signal-to-noise ratio in the observation
model~\eqref{EqnNetflix} is given by the ratio
\mbox{$\operatorname{SNR} =
  \frac{\Weifrosq{\ThetaStar}}{\noisevar^2}$.}

As shown by Negahban et al.~\cite{NegRavWaiYu09}, a key ingredient in
establishing error bounds for the observation model~\eqref{EqnNetflix}
is obtaining lower bounds on the restricted curvature of the sampling
operator---in particular, to establish the existence of a constant $c
> 0$, which may be arbitrarily small as long as it is positive, such
that
\begin{align}
\label{EqnPlainRSC}
\frac{\|\Xop(\Theta)\|_2}{\sqrt{\numobs}} & \geq c \: \Weifro{\Theta}.
\end{align}
For sample sizes of interest for matrix completion ($\numobs \ll \mrow
\mcol$) , one cannot expect such a bound to hold uniformly over all
matrices $\Theta \in \real^{\mrow \times \mcol}$, even when rank
constraints are imposed.  Indeed, as noted by Candes and
Plan~\cite{CanPla09}, the condition~\eqref{EqnPlainRSC} is violated
with high probability by the rank one matrix $\ThetaStar$ such that
$\ThetaStar_{11} = 1$ with all other entries zero.  Indeed, for a
sample size $\numobs \ll \mrow \mcol$, we have a vanishing probability
of observing the entry $\ThetaStar_{11}$, so that $\Xop(\ThetaStar) =
0$ with high probability.

\subsection{Controlling the spikiness and rank}

Intuitively, one must exclude matrices that are overly ``spiky'' in
order to avoid the phenomenon just described.  Past work has relied on
fairly restrictive matrix incoherence conditions (see
Section~\ref{SecCompare} for more discussion), based on specific
conditions on singular vectors of the unknown matrix $\ThetaStar$.  In
this paper, we formalize the notion of ``spikiness'' in a natural and
less restrictive way---namely by comparing a weighted form of
$\ell_\infty$-norm to the weighted Frobenius norm.  In particular, for
any non-zero matrix $\Theta$, let us define (for any non-zero matrix)
the \emph{weighted spikiness ratio}
\begin{align}
\label{EqnDefnSpike}
\alspike(\Theta) & \defn \MTIMES \; \; 
\frac{\Weiplinf{\Theta}}{\Weifro{\Theta}},
\end{align}
where $\Weiplinf{\Theta} \defn \|\sqrt{\Row} \Theta
\sqrt{\Col}\|_\infty$ is the weighted elementwise $\ell_\infty$-norm.
Note that this ratio is invariant to the scaling of $\Theta$, and
satisfies the inequalities $1 \leq \alspike(\Theta) \leq \MTIMES$.  We
have $\alspike(\Theta) = 1$ for any non-zero matrix whose entries are
all equal, whereas the opposite extreme \mbox{$\alspike(\Theta) =
  \MTIMES$} is achieved by the ``maximally spiky'' matrix that is zero
everywhere except for a single position.

In order to provide a tractable measure of how close $\Theta$ is to a
low-rank matrix, we define (for any non-zero matrix) the ratio
\begin{align}
\alrank(\Theta) & \defn \frac{\Weinuc{\Theta}}{\Weifro{\Theta}}
\end{align}
which satisfies the inequalities $1 \leq \alrank(\Theta) \leq
\sqrt{\min \{\mrow, \mcol\}}$.  By definition of the (weighted)
nuclear and Frobenius norms, note that $\alrank(\Theta)$ is simply the
ratio of the $\ell_1$ to $\ell_2$ norms of the singular values of the
weighted matrix $\sqrt{\Row} \Theta \sqrt{\Col}$.  This measure can
also be upper bounded by the rank of $\Theta$: indeed, since $\Row$
and $\Col$ are full-rank, we always have
\begin{align*}
\alrank^2(\Theta) & \leq \rank(\sqrt{\Row} \Theta \sqrt{\Col}) \; = \;
\rank(\Theta),
\end{align*}
with equality holding if all the non-zero singular values of
$\sqrt{\Row} \Theta \sqrt{\Col}$ are identical.


\section{Main results and their consequences}
\label{SecMain}

We now turn to the statement of our main results, and discussion of
their consequences.  Section~\ref{SecRSC} is devoted to a result
showing that a suitable form of restricted strong convexity holds for
the random sampling operator $\Xop$, as long as we restrict it to
matrices $\Delta$ for which $\alrank(\Delta)$ and $\alspike(\Delta)$
are not ``overly large''.  In Section~\ref{SecMatrixEst}, we develop
the consequences of the RSC condition for noisy matrix completion, and
in Section~\ref{SecLower}, we prove that our error bounds are
minimax-optimal up to logarithmic factors.  In
Section~\ref{SecCompare}, we provide a detailed comparison of our
results with past work.

\subsection{Restricted strong convexity for matrix sampling}
\label{SecRSC}

Introducing the convenient shorthand $\MSUM = \frac{1}{2}(\mrow +
\mcol)$, let us define the constraint set
\begin{align}
\label{EqnDefnConstraint}
\THMCON & \defn \biggr \{ \Delta \in
  \real^{\mrow \times \mcol}, \, \Delta \neq 0 \, \mid \,
  \alspike(\Delta) \; \alrank(\Delta) \leq \frac{1}{\Kcon}
  \sqrt{\frac{\numobs}{\MSUM \log \MSUM }} \, \biggr \},
\end{align}
where $\Kcon$ is a universal constant.  Note that as the sample size
$\numobs$ increases, this set allows for matrices with larger values
of the spikiness and/or rank measures, $\alspike(\Delta)$ and
$\alrank(\Delta)$ respectively.

\btheos
\label{ThmRSC}
There are universal constants $(\Kcon, \plaincon_1, \plaincon_2,
\plaincon_3)$ such that as long as $\numobs > \plaincon_3 \, \MSUM
\log \MSUM$, we have
\begin{align}
\label{EqnRSC}
\frac{\|\Xop(\Delta)\|_2}{\sqrt{\numobs}} & \geq \frac{1}{8}
\Weifro{\Delta} \; \biggr \{ 1 - \frac{128 \,
\alspike(\Delta)}{\sqrt{\numobs}} \biggr \} \qquad \mbox{for all
$\Delta \in \THMCON$}
\end{align}
with probability greater than $1 - \plaincon_1 \exp(-\plaincon_2 \MSUM
\log \MSUM)$.
\etheos

Roughly speaking, this bound guarantees that the observation operator
captures a substantial component of any matrix $ \Delta \in \THMCON$
that is not overly spiky.  More precisely, as long as
$\frac{128 \alspike(\Delta)}{\sqrt{\numobs}} \leq \frac{1}{2}$, the
bound~\eqref{EqnRSC} implies that
\begin{align}
\label{EqnConcrete}
\frac{\|\Xop(\Delta)\|^2_2}{\numobs} & \geq \frac{1}{256}
\Weifrosq{\Delta} \qquad \mbox{for any $\Delta \in \THMCON$.}
\end{align}
This bound can be interpreted in terms of \emph{restricted strong
convexity}~\cite{NegRavWaiYu09}.  In particular, given a vector $y \in
\real^\numobs$ of noisy observations, consider the quadratic loss
function
\begin{align*}
\mathcal{L}(\Theta; y) & = \frac{1}{2 \numobs} \|y - \Xop(\Theta)\|_2^2.
\end{align*}
Since the Hessian matrix of this function is given by $\Xop^*
\Xop/\numobs$, the bound~\eqref{EqnConcrete} implies that the
quadratic loss is strongly convex in a restricted set of directions
$\Delta$.

As discussed previously, the worst-case value of the ``spikiness''
measure is \mbox{$\alspike(\Delta) = \sqrt{\mrow \mcol}$,} achieved
for a matrix that is zero everywhere except a single position.  In
this most degenerate of cases, the combination of the constraints
$\frac{\alspike(\Delta)}{\sqrt{\numobs}} < 1$ and the membership
condition $\Delta \in \THMCON$ imply that even for a rank one matrix
(so that $\alrank(\Delta) = 1$), we need sample size $\numobs \gg
\mdim^2$ for Theorem~\ref{ThmRSC} to provide a non-trivial result, as
is to be expected. \\


\subsection{Consequences for noisy matrix completion}
\label{SecMatrixEst}
We now turn to some consequences of Theorem~\ref{ThmRSC} for matrix
completion in the noisy setting.  In particular, assume that we are
given $\numobs$ i.i.d. samples from the model~\eqref{EqnNetflix}, and
let $\ThetaHat$ be some estimate of the unknown matrix $\ThetaStar$.
Our strategy is to exploit the lower bound~\eqref{EqnRSC} in
application to the error matrix $\ThetaHat - \ThetaStar$, and
accordingly, we need to ensure that it has relatively low-rank and
spikiness.  Based on this intuition, it is natural to consider the
estimator
\begin{align}
\label{EqnWeightSDP}
\ThetaHat & \in \arg \min_{\Weiplinf{\Theta} \leq
\frac{\spikeupper}{\MTIMES}} \big \{ \frac{1}{2 \numobs} \|y -
\Xop(\Theta)\|_2^2 + \regpar \Weinuc{\Theta} \big \},
\end{align}
where $\spikeupper \geq 1$ is a measure of spikiness, and the
regularization parameter $\regpar > 0$ serves to control the nuclear
norm of the solution.  In the special case when both $\Row$ and $\Col$
are identity matrices (of the appropriate dimensions), this estimator
is closely related to the standard one considered in past work on the
problem, with the only difference between the additional
\mbox{$\ell_\infty$-norm} constraint.  In the more general weighted
case, an $M$-estimator of the form~\eqref{EqnWeightSDP} using the
weighted nuclear norm (but without the elementwise constraint) was
recently suggested by Salakhutdinov and Srebro~\cite{SalSre10}, who
provided empirical results to show superiority of the weighted nuclear
norm over the standard choice for the Netflix problem.

Past work on matrix completion has focused on the case of exactly
low-rank matrices.  Here we consider the more general setting of
approximately low-rank matrices, including the exact setting as a
particular case.  We begin by stating a general upper bound that
applies to any matrix $\ThetaStar$, and involves a natural
decomposition into estimation and approximation error terms.
\btheos
\label{ThmMatrixEst}
Consider any solution $\ThetaHat$ to the weighted
SDP~\eqref{EqnWeightSDP} using regularization parameter
\begin{align}
\label{EqnRegChoice}
\regpar & \geq 2 \noisevar \, \opnorm{\frac{1}{\numobs}
\sum_{i=1}^\numobs \newvsca_i \Row^{-\frac{1}{2}} \ObsMat{i}
\Col^{-\frac{1}{2}}},
\end{align}
and define $\regpar^* = \max \{ \regpar, \sqrt{\frac{\mdim \log
    \mdim}{\numobs}} \}$.  Then with probability greater than $1 -
\plaincon_2 \exp(-\plaincon_2 \log \mdim)$, for each $\rdim = 1,
\ldots, \mrow$, the error $\DeltaTil = \ThetaHat - \ThetaStar$
satisfies
\begin{align}
\label{EqnMatrixEst}
\Weifrosq{\DeltaTil} & \leq \plaincon_1 \; \upperspike \; \regpar^* \;
\; \biggr [ \sqrt{\rdim} \Weifro{\DeltaTil} + \sum_{j=\rdim+1}^{\mrow}
\sigma_j(\sqrt{\Row} \ThetaStar \sqrt{\Col}) \biggr].
\end{align}
\etheos

Notice how the bound~\eqref{EqnMatrixEst} shows a natural splitting
into two terms.  The first can be interpreted as the \emph{estimation
error} associated with a rank $\rdim$ matrix, whereas the second term
corresponds to \emph{approximation error}, measuring how far
$\sqrt{\Row} \ThetaStar \sqrt{\Col}$ is from a rank $\rdim$ matrix.
Of course, the bound holds for any choice of $\rdim$, and in the
corollaries to follow, we choose $\rdim$ optimally so as to balance
the estimation and approximation error terms. \\

In order to provide concrete rates using Theorem~\ref{ThmMatrixEst},
it remains to address two issues.  First, we need to specify an
explicit choice of $\regpar$ by bounding the operator norm of the
matrix $\frac{1}{\numobs} \sum_{i=1}^\numobs \newvsca_i \sqrt{\Row}
\ObsMat{i} \sqrt{\Col}$, and secondly, we need to understand how to
choose the parameter $\rdim$ so as to achieve the tightest possible
bound.  When $\ThetaStar$ is exactly low-rank, then it is obvious that
we should choose $\rdim = \rank(\ThetaStar)$, so that the
approximation error vanishes---viz.  $\sum_{j=\rdim+1}^\mrow
\sigma_j(\sqrt{\Row} \ThetaStar \sqrt{\Col})_j = 0$. Doing so yields
the following result:

\bcors[Exactly low-rank matrices]
\label{CorExact}
Suppose that the noise sequence $\{\newvsca_i\}$ is i.i.d., zero-mean
and sub-exponential, and $\ThetaStar$ has rank at most $\rdim$,
Frobenius norm at most $1$, and spikiness at most
$\alspike(\ThetaStar) \leq \upperspike$.  If we solve the
SDP~\eqref{EqnWeightSDP} with $\regpar = 4 \Lcon \noisevar
\sqrt{\frac{\mdim \log \mdim}{\numobs}}$ then there is a numerical
constant $\plaincon'_1$ such that
\begin{align}
\label{EqnExact}
\Weifrosq{\ThetaHat - \ThetaStar} & \leq \plaincon'_1 \; (\noisevar^2
\vmax 1) \; (\upperspike)^2 \; \: \frac{\rdim \mdim \log
\mdim}{\numobs}
\end{align}
with probability greater than $1 - \plaincon_2 \exp(-\plaincon_3 \log
\mdim)$.
\ecors

\noindent Note that this rate has a natural interpretation: since a
rank $\rdim$ matrix of dimension $\mrow \times \mcol$ has roughly
$\rdim (\mrow + \mcol)$ free parameters, we require a sample size of
this order (up to logarithmic factors) so as to obtain a controlled
error bound.  An interesting feature of the bound~\eqref{EqnExact} is
the term $\noisevar^2 \vmax 1 = \max \{ \noisevar^2, 1\}$, which
implies that we do not obtain exact recovery as $\noisevar \rightarrow
0$.  As we discuss at more length in Section~\ref{SecCompare}, under
the mild spikiness condition that we have imposed, this behavior is
unavoidable due to lack of identifiability within a certain radius, as
specified in the set $\Constraint$. For instance, consider the matrix
$\ThetaStar$ and the perturbed version $\ThetaTil = \ThetaStar +
\frac{1}{\MTIMES} e_1 e_1^T$.  With high probability, we have
$\Xop(\ThetaStar) = \Xop(\ThetaTil)$, so that the observations---even
if they were noiseless---fail to distinguish between these two models.
These types of examples, leading to non-identifiability, cannot be
overcome without imposing fairly restrictive matrix incoherence
conditions, as we discuss at more length in Section~\ref{SecCompare}.

\vspace*{.1in}

As with past work~\cite{CanPla09,Keshavan09b},
Corollary~\ref{CorExact} applies to the case of matrices that have
exactly rank $\rdim$.  In practical settings, it is more realistic to
assume that the unknown matrix is not exactly low-rank, but rather can
be well approximated by a matrix with low rank.  One way in which to
formalize this notion is via the $\ell_\qpar$-``ball'' of matrices
\begin{align}
\label{EqnDefnQparBall}
\Ball_\qpar(\radq) & \defn \biggr \{ \Theta \in \real^{\mrow \times
  \mcol} \, \mid \, \sum_{j=1}^{\min\{\mrow, \mcol\}}
  |\sigma_j(\sqrt{\Row} \Theta \sqrt{\Col})|^\qpar \leq \radq \biggr
  \}.
\end{align}
For $\qpar = 0$, this set corresponds to the set of matrices with rank
at most $\rdim = \radplain_0$, whereas for values $\qpar \in (0,1]$,
  it consists of matrices whose (weighted) singular values decay at a
  relatively fast rate.  By applying Theorem~\ref{ThmMatrixEst} to
  this matrix family, we obtain the following corollary:
\bcors[Estimation of near low-rank matrices]
\label{CorApprox}
Suppose that the noise $\{\newvsca_i\}$ is zero-mean and sub-exponential, and
$\ThetaStar \in \Ball_\qpar(\radq)$ and has spikiness at most
$\alspike(\ThetaStar) \leq \upperspike$.  With the same choice of
$\regpar$ as Corollary~\ref{CorExact}, there is a universal constant
$\plaincon'_1$ such that
\begin{align}
\label{EqnApproxBound}
\Weifrosq{\ThetaHat - \ThetaStar} & \leq \plaincon_1 \radq \; \Big(
(\noisevar^2 \vmax 1) (\upperspike)^2 \frac{\mdim \log \mdim}{\numobs}
\Big)^{1 - \frac{\qpar}{2}}
\end{align}
with probability greater than $1 - \plaincon_2 \exp(-\plaincon_3 \log
\mdim)$.
\ecors
Note that this result is a strict generalization of
Corollary~\ref{CorExact}, to which it reduces in the case $\qpar =
0$. (When $\qpar = 0$, we have $\radplain_0 = \rdim$ so that the bound
has the same form.)  Note that the price that we pay for approximately
low rank is a smaller exponent---namely, $1-\qpar/2$ as opposed to $1$
in the case $\qpar = 0$.  The proof of Corollary~\ref{CorApprox} is
based on a more subtle application of Theorem~\ref{ThmMatrixEst}, one
which chooses the effective rank $\rdim$ in the
bound~\eqref{EqnMatrixEst} so as to trade off between the estimation
and approximation errors.  In particular, the choice $\rdim \asymp
\radq \, (\frac{\numobs}{\mdim \log \mdim})^{\qpar/2}$ turns out to
yield the optimal trade-off, and hence the given error
bound~\eqref{EqnApproxBound}. \\

\vspace*{.05in}

In order to illustrate the sharpness of our theory, let us compare the
predictions of our two corollaries to the empirical behavior of the
$M$-estimator.  In particular, we applied the nuclear norm SDP to
simulated data, using Gaussian observation noise with variance
$\noisevar^2 = 0.25$ and the uniform sampling model. In all cases, we
solved the nuclear norm SDP using a non-smooth optimization procedure
due to Nesterov~\cite{Nesterov07}, via our own implementation in
MATLAB.  For a given problem size $\mdim$, we ran $T = 25$ trials and
computed the squared Frobenius norm error $\matsnorm{\ThetaHat -
\ThetaStar}{F}^2$ averaged over the trials.

\begin{figure}[h]
\begin{center}
\begin{tabular}{cc}
\widgraph{.45\textwidth}{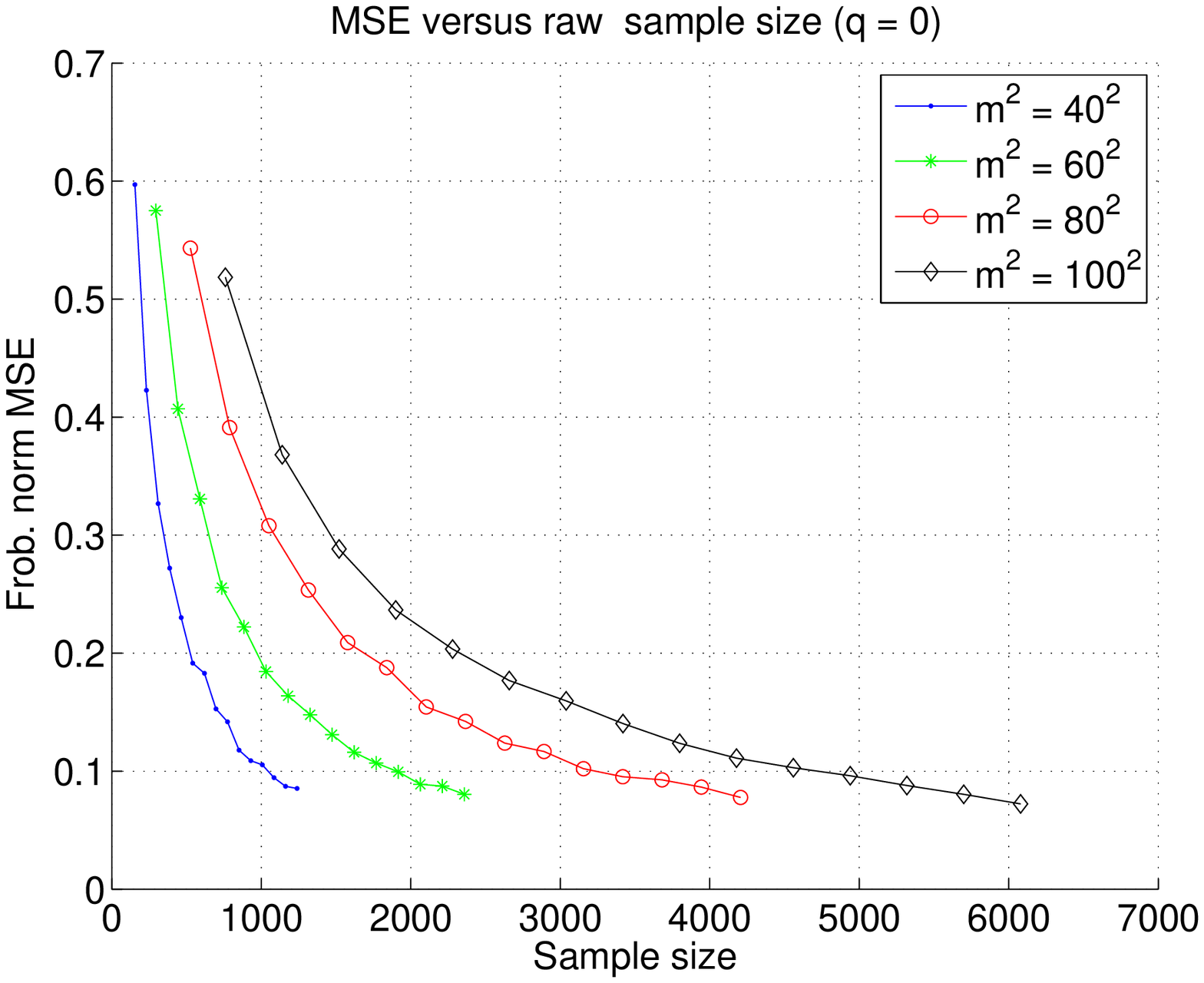} &
\widgraph{.44\textwidth}{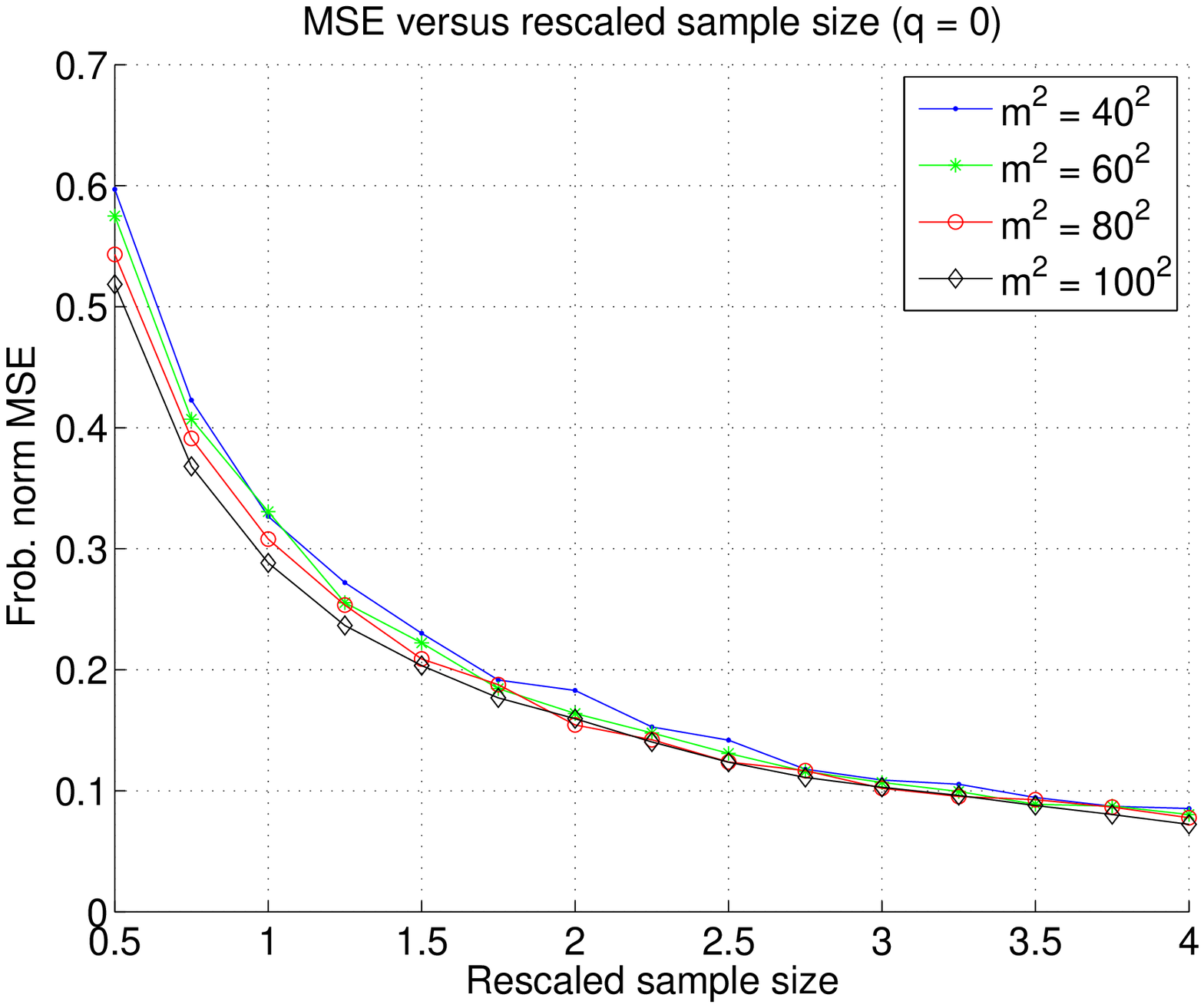} \\
(a) & (b)
\end{tabular}
\end{center}
\caption{Plots of the mean-squared error in Frobenius norm for $\qpar
= 0$.  Each curve corresponds to a different problem size 
$\mdim^2 \in
\{40^2, 60^2, 80^2, 100^2 \}$.  (a) MSE versus the raw sample size
$\numobs$.  As expected, the curves shift to the right as $\mdim$
increases, since more samples should be required to achieve a given
MSE for larger problems.  (b) The same MSE plotted versus the rescaled
sample size $\numobs/(\rdim \mdim \log \mdim)$.  Consistent with
Corollary~\ref{CorExact}, all the plots are now fairly well-aligned.}
\label{FigSimulationsQzero}
\end{figure}

Figure~\ref{FigSimulationsQzero} shows the results in the case of
exactly low-rank matrices ($\qpar = 0$), with the matrix rank given by
$r = \lceil \log^2(\mdim) \rceil$.  Panel (a) shows plots of the
mean-squared Frobenius error versus the raw sample size, for three
different problem sizes with the number of matrix elements sizes
$\mdim^2 \in \{40^2, 60^2, 80^2, 100^2 \}$.  These plots show that the
$M$-estimator is consistent, since each of the curves decreases to
zero as the sample size $\numobs$ increases.  Note that the curves
shift to the right as the matrix dimension $\mdim$ increases,
reflecting the natural intuition that larger matrices require more
samples.  Based on the scaling predicted by Corollary~\ref{CorExact},
we expect that the mean-squared
Frobenius error should exhibit the scaling $\matsnorm{\ThetaHat -
\ThetaStar}{F}^2 \asymp \frac{\rdim \mdim \log \mdim}{\numobs}$.
Equivalently, if we plot the MSE versus the \emph{rescaled sample
size} $N \defn \frac{\numobs}{\rdim \mdim \log \mdim}$, then all the
curves should be relatively well aligned, and decay at the rate $1/N$.
Panel (b) of Figure~\ref{FigSimulationsQzero} shows the same
simulation results re-plotted versus this rescaled sample size.
Consistent with the prediction of Corollary~\ref{CorExact}, all four
plots are now relatively well-aligned.
\begin{figure}[h!]
\begin{center}
\begin{tabular}{cc}
 \widgraph{.45\textwidth}{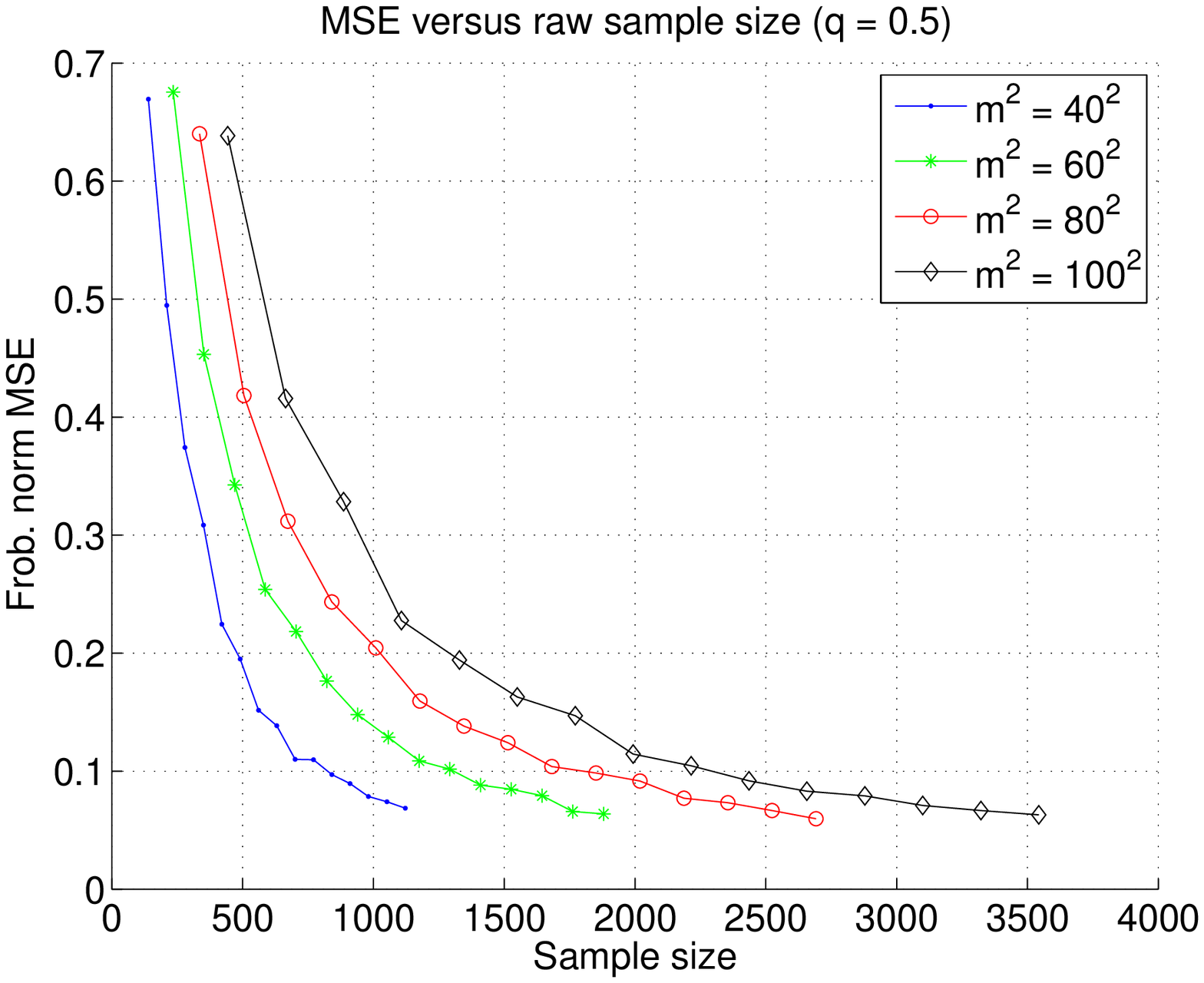} &
\widgraph{.44\textwidth}{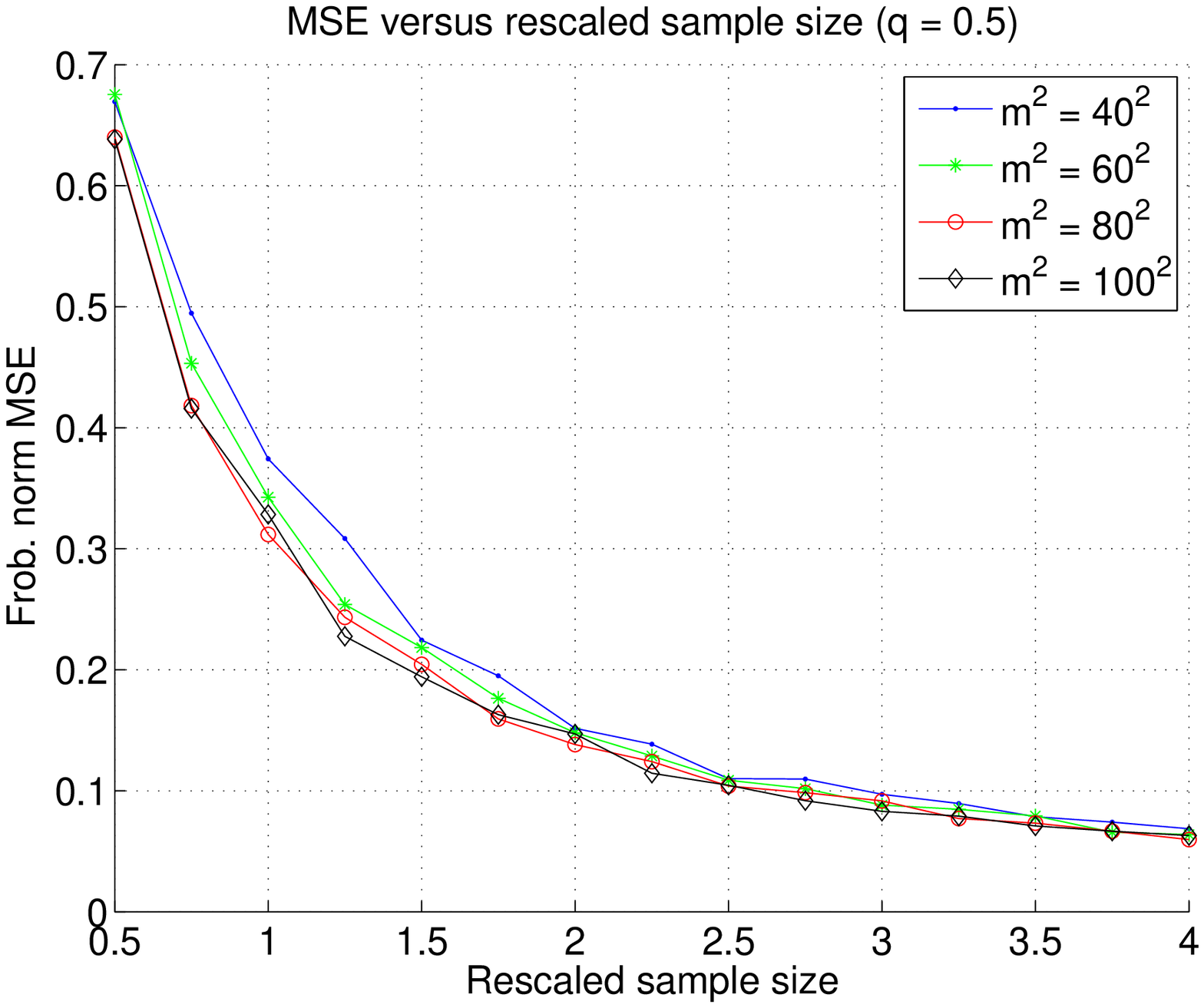} \\
(a) & (b)
\end{tabular}
\end{center}
\caption{Plots of the mean-squared error in Frobenius norm for $\qpar
= 0.5$.  Each curve corresponds to a different problem size $\mdim^2 \in
\{40^2, 60^2, 80^2, 100^2 \}$.  (a) MSE versus the raw sample size
$\numobs$.  As expected, the curves shift to the right as $\mdim$
increases, since more samples should be required to achieve a given
MSE for larger problems.  (b) The same MSE plotted versus the rescaled
sample size $\numobs/(\radq^{\frac{1}{1 - \qpar/2}} \, \mdim \log
\mdim)$.  Consistent with Corollary~\ref{CorApprox}, all the plots are
now fairly well-aligned.}
\label{FigSimulationsQ05}
\end{figure}
Figure~\ref{FigSimulationsQ05} shows the same plots for the case
of approximately low-rank matrices ($\qpar = 0.5$).  Again, consistent
with the prediction of Corollary~\ref{CorApprox}, we see qualitatively
similar behavior in the plots of the MSE versus sample size (panel (a)),
and the rescaled sample size (panel (b)).

\subsection{Information-theoretic lower bounds}
\label{SecLower}

The results of the previous section are achievable results, based on a
particular polynomial-time estimator.  It is natural to ask how these
bounds compare to the fundamental limits of the problem, meaning the
best performance achievable by any algorithm.  As various authors have
noted~\cite{CanPla09,Keshavan09b}, a parameter counting argument
indicates that roughly $\numobs \approx \rdim \, (\mrow + \mcol)$
samples are required to estimate an $\mrow \times \mcol$ matrix with
rank $\rdim$.  This calculation can be made more formal by metric
entropy calculations for the Grassman manifold (e.g.,~\cite{Sza83});
see also Rohde and Tsybakov~\cite{RohTsy10} for results on
approximation numbers for the more general $\ell_\qpar$-balls of
matrices.  Such calculations, while accounting for the low-rank
conditions, do \emph{not} address the additional ``spikiness''
constraints that are essential to the setting of matrix completion.
It is conceivable that these additional constraints could lead to a
substantial volume reduction in the allowable class of matrices, so
that the scalings suggested by parameter counting or metric entropy
calculation for Grassman manifolds would be overly conservative.

Accordingly, in this section, we provide a direct and constructive
argument to lower bound the minimax rates of Frobenius norm over
classes of matrices that are near low-rank and not overly spiky.  This
argument establishes that the bounds established in
Corollaries~\ref{CorExact} and~\ref{CorApprox} are sharp up to
logarithmic factors, meaning that no estimator performs substantially
better than the one considered here.  More precisely, consider the
matrix classes
\begin{align}
\Ballspec(\radq) & = \biggr \{ \Theta \in \real^{\mdim \times \mdim} \,
\mid \, \sum_{j=1}^\mdim \sigma_j(\Theta)^\qpar \leq \radq, \;
\alspike(\Theta) \leq \sqrt{32 \log \mdim} \biggr \},
\end{align}
corresponding to square $\mdim \times \mdim$ matrices that are near
low-rank (belonging to the $\ell_\qpar$-balls previously
defined~\eqref{EqnDefnQparBall}), and have a logarithmic spikiness
ratio.  The following result applies to the \emph{minimax risk} in
Frobenius norm, namely the quantity
\begin{align}
\MiniMax_\numobs(\Ballspec(\radq)) & \defn \inf_{\ThetaTil}
\sup_{\ThetaStar \in \Ballspec(\radq)} \Exs \big[ \matsnorm{\ThetaTil
- \ThetaStar}{F}^2 \big],
\end{align}
where the infimum is taken over all estimators $\ThetaTil$ that
are measurable functions of $\numobs$ samples.
\btheos
\label{ThmLower}
There is a universal numerical constant $\japan > 0$ such that 
\begin{align}
\MiniMax_\numobs(\Ballspec(\radq)) & \geq \japan \; \min \biggr \{
\radq \, \biggr(\frac{\noisevar^2 \mdim}{\numobs}
\biggr)^{1-\frac{\qpar}{2}}, 
\; \frac{\noisevar^2 \mdim^2}{\numobs}
\biggr \}.
\end{align}
\etheos
The term of primary interest in this bound is the first one---namely,
$\radq \, \big(\frac{\noisevar^2 \mdim}{\numobs}
\big)^{1-\frac{\qpar}{2}}$.  It is the dominant term in the bound
whenever the $\ell_\qpar$-radius satisfies the bound
\begin{align}
\label{EqnKeyBound}
\radq & \leq \biggr( \frac{\noisevar^2 \mdim}{\numobs}
\biggr)^{\frac{\qpar}{2}} \, \mdim.
\end{align}
In the special case $\qpar = 0$, corresponding the exactly low-rank
case, the bound~\eqref{EqnKeyBound} always holds, since it reduces to
requiring that the rank $\rdim = \radplain_0$ is less than or equal to
$\mdim$.  In these regimes, Theorem~\ref{ThmLower} establishes that
the upper bounds obtained in Corollaries~\ref{CorExact}
and~\ref{CorApprox} are minimax-optimal up to factors logarithmic in
matrix dimension $\mdim$.

\subsection{Comparison to other work}
\label{SecCompare}

We now turn to a detailed comparison of our bounds to those obtained
in past work on noisy matrix completion, in particular the papers by
Candes and Plan~\cite{CanPla09} (hereafter CP) and Keshavan et
al.~\cite{Keshavan09b} (hereafter KMO).  Both papers considered only
the case of exactly low-rank matrices, corresponding to the special
case of $\qpar = 0$ in our notation.  Since neither paper provided
results for the general case of near-low rank matrices, nor the
general result (with estimation and approximation errors) stated in
Theorem~\ref{ThmMatrixEst}, our discussion is limited to comparing
Corollary~\ref{CorExact} to their results.  So as to simplify
discussion, we restate all results under the scalings used in this
paper\footnote{The paper CP and KMO use two different sets of scaling,
  one with $\matsnorm{\ThetaStar}{F} = \Theta(\mdim)$ and the other
  with $\matsnorm{\ThetaStar}{F} = \sqrt{\rdim}$, so that some care is
  required in converting between results.} (i.e., with
$\matsnorm{\ThetaStar}{F} = 1$).

\subsubsection{Comparison of rates}
Under the strong incoherence conditions required for exact matrix
recovery (see below for discussion), Theorem~{7} in CP give an bound
on $\matsnorm{\ThetaHat - \ThetaStar}{F}$ that depends on the
Frobenius norm of the error matrix $\Xi \in \real^{\mdima \times
  \mdimb}$, as defined by the noise variables $[\Xi]_{j(i) \; k(i)} =
\vscatil_{i}$ in our case.  Under the observation
model~\eqref{EqnOrigObs} and the scalings of our paper, as long as
$\numobs > \mdim$, where $\mdim = \mdima + \mdimb$---a condition
certainly required for Frobenius norm consistency---we have
$\matsnorm{\Xi}{F} = \Theta( \noisevar \sqrt{\numobs}/\mdim)$ with high
probability.  Given this scaling, the CP upper bound takes the form
\begin{align}
\label{EqnCPTwo}
\matsnorm{\ThetaHat - \ThetaStar}{F} & \lesssim \noisevar \, \biggr \{
\sqrt{\mdim} + \frac{\sqrt{\numobs}}{\mdim} \biggr \}.
\end{align}
Note that if the noise standard deviation $\noisevar$ tends to zero
while the sample size $\numobs$, matrix size $\pdim$ and rank $\rdim$
all remain fixed, then this bound guarantees that the Frobenius error
tends to zero.  This behavior as $\noisevar \rightarrow 0$ is
intuitively reasonable, given that their proof technique is an
extrapolation from the case of exact recovery for noiseless
observations ($\noisevar = 0$).  However, note that for any fixed
noise deviation $\noisevar > 0$, the first term increases to infinity
as the matrix dimension $\mdim$ increases, whereas the second term
actually grows as the sample size $\numobs$ increases.  Consequently,
the CP results do not guarantee statistical consistency, unlike the
bounds proved here.

Keshavan et al.~\cite{Keshavan09b} analyzed alternative methods based
on trimming and applying the SVD.  For Gaussian noise, their methods
guarantee bounds (with high probability) of the form
\begin{align}
\label{EqnKMOBound}
\frobnorm{\ThetaHat - \ThetaStar} & \lesssim \noisevar \min \big \{
\alpha \, \sqrt{\frac{\mdimb}{\mdima}}, \kappa^2(\ThetaStar) \big \}
\; \sqrt{\frac{\rdim \mdimb }{\numobs}},
\end{align}
where $\mdimb/\mdima$ is the aspect ratio of $\ThetaStar$, and
$\kappa(\ThetaStar) = \frac{\sigmax(\ThetaStar)}{\sigmin(\ThetaStar)}$
is the condition number of $\ThetaStar$.  This result is more directly
comparable to our Corollary~\ref{CorExact}; apart from the additional
factor involving either the aspect ratio or the condition number, it
is sharper since it does not involve the factor $\log \mdim$ present
in our bound.  For a fixed noise standard deviation $\noisevar$, the
bound~\eqref{EqnKMOBound} guarantees statistical consistency as long
as $\frac{\rdim \mdimb}{\numobs}$ tends to zero.  The most significant
differences are the presence of the aspect ratio $\mdimb/\mdima$ or
the condition number $\kappa(\ThetaStar)$ in the upper
bound~\eqref{EqnKMOBound}.  The aspect ratio is a quantity that can be
as small as one, or as large as $\mdimb$, so that the pre-factor in
the bound~\eqref{EqnKMOBound} can scale in a dimension-dependent way.
Similarly, for any matrix with rank larger than one, the condition
number can be made arbitrarily large.  For instance, in the rank two
case, define a matrix with $\sigmax(\ThetaStar) = \sqrt{1 - \delta^2}$
and $\sigmin(\ThetaStar) = \delta$, and consider the behavior as
$\delta \rightarrow 0$.  In contrast, our bounds are invariant to both
the aspect ratio and the condition number of $\ThetaStar$.

\subsubsection{Comparison of matrix conditions}

We now turn to a comparison of the various \emph{matrix incoherence
  assumptions} invoked in the analysis of CP and KMO, and comparison
to our spikiness condition.  As before, for clarity, we specialize our
discussion to the square case ($\mrow = \mcol = \mdim$), since the
rectangular case is not essentially different. The matrix incoherence
conditions are stated in terms of the singular value decomposition
$\ThetaStar = U \Sigma V^T$ of the target matrix. Here $U \in
\real^{\mdim \times \rdim}$ and $V \in \real^{\mdim \times \rdim}$ are
matrices of the left and right singular vectors respectively,
satisfying \mbox{$U^T U = V^T V = I_{\rdim \times \rdim}$,} whereas
$\Sigma \in \real^{\rdim \times \rdim}$ is a diagonal matrix of the
singular values.  The purpose of matrix incoherence is to enforce that
the left and right singular vectors should not be aligned with the
standard basis.  Among other assumptions, the CP analysis imposes the
incoherence conditions
\begin{align}
\label{CanPlaAss1} 
\|U U^T - \frac{\rdim}{\mdim} I_{\mdim \times \mdim} \|_\infty \leq
\mu \, \frac{\sqrt{\rdim}}{\mdim}, \qquad \|V V^T -
\frac{\rdim}{\mdim} I_{\mdim \times \mdim} \|_\infty \leq \mu \,
\frac{\sqrt{\rdim}}{\mdim}, \quad \mbox{and} \quad
\| U V^T \|_\infty \leq \mu \, \frac{\sqrt{\rdim}}{\mdim},
\end{align}
for some constant $\mu > 0$.  Parts of the KMO analysis impose the
related incoherence condition
\begin{equation}
\label{KesMonOhAss1}
\max_{j = 1, \ldots, \mdim} |U U^T|_{jj} \leq \; \mu_0
\frac{\rdim}{\mdim}, \quad \mbox{and} \max_{j = 1, \ldots, \mdim} |V
V^T|_{jj} \; \leq \; \mu_0 \, \frac{\rdim}{\mdim}.
\end{equation}
Both of these conditions ensure that the singular vectors are
sufficiently ``spread-out'', so as not to be aligned with the standard
basis.

A remarkable property of conditions~\eqref{CanPlaAss1}
and~\eqref{KesMonOhAss1} is that they exhibit \emph{no dependence} on
the singular values of $\ThetaStar$.  If one is interested only in
exact recovery in the noiseless setting, then this lack of dependence
is reasonable.  However, if approximate recovery is the goal---as is
necessarily the case in the more realistic setting of noisy
observations---then it is clear that a minimal set of sufficient
conditions should also involve the singular values, as is the case for
our spikiness measure $\alspike(\ThetaStar)$.  The following example
gives a concrete demonstration of an instance where our conditions are
satisfied, so that approximate recovery is possible, whereas the
incoherence conditions are violated. \\

%
%
\noindent {\bf{Example.}}
Let $\Gamma \in \real^{\mdim \times \mdim}$ be a positive semidefinite
symmetric matrix with rank $\rdim-1$, Frobenius norm
$\matsnorm{\Gamma}{F} = 1$ and $\|\Gamma\|_\infty \leq c_0/\mdim$.
For a scalar parameter $t > 0$, consider the matrix
\begin{align}
\ThetaStar & \defn \Gamma +  t e_1 e_1^T
\end{align}
where $e_1 \in \real^\mdim$ is the canonical basis vector with one in
its first entry, and zero elsewhere.  By construction, the matrix
$\ThetaStar$ has rank at most $\rdim$.  Moreover, as long as $t =
\order(1/\mdim)$, we are guaranteed that our spikiness measure
satisfies the bound $\alspike(\ThetaStar) = \order(1)$.  Indeed, we
have $\matsnorm{\ThetaStar}{F} \geq \matsnorm{\Gamma}{F} - t \; = 1
-t$, and hence
\begin{align*}
\alspike(\ThetaStar) & = \frac{\mdim
  \|\ThetaStar\|_\infty}{\matsnorm{\ThetaStar}{F}} \; \leq \; \frac{
  \mdim \big( \|\Gamma\|_\infty + t \big)}{1-t} \; \leq \; \frac{c_0 +
  \mdim t}{1 -t} \; = \; \order(1).
\end{align*}
Consequently, for any choice of $\Gamma$ as specified above,
Corollary~\ref{CorExact} implies that the SDP will recover the matrix
$\ThetaStar$ up to a tolerance $\order(\sqrt{\frac{\rdim \mdim \log
\mdim}{\numobs}})$.  This captures the natural intuition that
``poisoning'' the matrix $\Gamma$ with the term $t e_1^T e_1$ should
have essentially no effect, as long as $t$ is not too large.

On the other hand, suppose that we choose the matrix $\Gamma$ such
that its $\rdim-1$ eigenvectors are orthogonal to $e_1$.  In this
case, we have $\ThetaStar e_1 = t e_1$, so that $e_1$ is also an
eigenvector of $\ThetaStar$.  Letting $U \in \real^{\mdim \times
\rdim}$ be the matrix of eigenvectors, we have $e_1^T U U^T e_1 = 1$.
Consequently, for any fixed $\mu$ (or $\mu_0$) and rank $\rdim \ll
\mdim$, conditions~\eqref{CanPlaAss1} and~\eqref{KesMonOhAss1} are
violated.

\hfill $\diamondsuit$
%
%


\section{Proofs for noisy matrix completion}
\label{SecProofThmMatrixEst}

We now turn to the proofs of our results.  This section is devoted to
the results that apply directly to noisy matrix completion, in
particular the achievable result given in Theorem~\ref{ThmMatrixEst},
its associated Corollaries~\ref{CorExact} and~\ref{CorApprox}, and the
information-theoretic lower bound given in Theorem~\ref{ThmLower}.
The proof of Theorem~\ref{ThmRSC} is provided in
Section~\ref{SecProofThmRSC} to follow.

\subsection{A useful transformation}

We begin by describing a transformation that is useful both in these
proofs, and the later proof of Theorem~\ref{ThmRSC}.  In particular, we
consider the mapping $\Theta \mapsto \Gamma \defn \sqrt{\Row} \Theta
\sqrt{\Col}$, as well as the modified observation operator $\XopNew:
\real^{\prow \times \pcol} \rightarrow \real^\numobs$ with elements
\begin{align*}
[\XopNew(\GamPlain)]_i & = \tracer{\ObsTil{i}}{\GamPlain}, \quad
\mbox{for $i = 1, 2, \ldots, \numobs$,}
\end{align*}
where $\ObsTil{i} \defn \Row^{-1/2} \, \ObsMat{i} \Col^{-1/2}$.  Note
that $\XopNew(\Gamma) = \Xop(\Theta)$ by construction, and moreover
\begin{equation*}
\matsnorm{\GamPlain}{F} \; = \; \Weifro{\Theta}, \quad
\matsnorm{\GamPlain}{1} \; = \; \Weinuc{\Theta}, \quad \mbox{and}
\quad \matsnorm{\GamPlain}{\infty} \; = \; \Weiplinf{\Theta},
\end{equation*}
which implies that
\begin{equation}
\alrank(\Theta) \, = \,
\underbrace{\frac{\matsnorm{\GamPlain}{1}}{\matsnorm{\GamPlain}{F}}}_{
\altworank(\GamPlain)} , \quad \mbox{and} \quad
\alspike(\Theta) \, = \, \underbrace{\frac{\MTIMESPOST \;
\|\GamPlain\|_\infty}{\matsnorm{\GamPlain}{F}}}_{\altwospike(\GamPlain)}.
\end{equation}
Based on this change of variables, let us define a modified version of
the constraint set~\eqref{EqnDefnConstraint} as follows
\begin{align}
\label{EqnDefnConstraintNew}
\THMCONNEW & = \biggr \{ 0 \neq \GamPlain \in \real^{\prow \times
 \pcol} \, \mid \, \altwospike(\GamPlain) \; \altworank(\GamPlain)
 \leq \frac{1}{\Kcon} \sqrt{\frac{\numobs}{\MSUM \log \MSUM}} \,
 \biggr \}.
\end{align}
In this new notation, the lower bound~\eqref{EqnRSC} from
Theorem~\ref{ThmRSC} can be re-stated as
\begin{align}
\label{EqnRSCNew}
\frac{\|\XopNew(\Gamma)\|_2}{\sqrt{\numobs}} & \geq \frac{1}{8}
\matsnorm{\Gamma}{F} \big \{ 1 - \frac{128
\altwospike(\Gamma)}{\sqrt{\numobs}} \big \} \quad \mbox{for all
$\Gamma \in \THMCONNEW$.}
\end{align}

\subsection{Proof of Theorem~\ref{ThmMatrixEst}}

We now turn to the proof of Theorem~\ref{ThmMatrixEst}.  Defining the
estimate \mbox{$\GammaHat \defn \sqrt{\Row} \ThetaHat \sqrt{\Col}$,}
we have
\begin{align}
\label{EqnGammaSDP}
\GammaHat & \in \arg \min_{\|\Gamma\|_\infty \leq
\frac{\spikeupper}{\MTIMES}} \big \{ \frac{1}{2 \numobs} \|y -
\XopNew(\Gamma)\|_2^2 + \regpar \matsnorm{\Gamma}{1} \big \},
\end{align}
and our goal is to upper bound the ordinary Frobenius norm
$\matsnorm{\GammaHat - \GammaStar}{F}$.  

We now state a useful technical result.  Parts (a) and (b) of the
following lemma were proven by Recht et al.~\cite{RecFazPar07} and
Negahban and Wainwright~\cite{NegWai09}, respectively.

\blems
\label{LemRestricted}
Let $(\mataspec, \matbspec)$ represent a pair of $\rdim$-dimensional
subspaces of left and right singular vectors of $\GammaStar$.  Then
there exists a matrix decomposition $\PlError = \PlErrorA + \PlErrorB$
of the error $\PlError$ such that
\begin{enumerate}
\item[(a)] The matrix $\PlErrorA$ satisfies the constraint
$\rank(\PlErrorA) \leq 2 \rdim$, and
\item[(b)] Given the choice~\eqref{EqnRegChoice}, the nuclear norm of
 $\PlErrorB$ is bounded as
\begin{equation}
\label{EqnRestricted}
  \nucnorm{\PlErrorB} \; \leq \; 3 \nucnorm{\PlErrorA} \, + \, 4
\TAILSUM.
\end{equation}
\end{enumerate}
\elems
Note that the bound~\eqref{EqnRestricted}, combined with triangle
inequality, implies that
\begin{align}
\nucnorm{\DeltaHat} & \leq \nucnorm{\PlErrorA} + \nucnorm{\PlErrorB}
  \; \leq \; 4 \nucnorm{\PlErrorA} \, + \, 4 \TAILSUM \nonumber \\
\label{EqnRestrictedTwo}
& \leq 8 \sqrt{\rdim} \matsnorm{\DeltaHat}{F} + 4 \TAILSUM
\end{align}
where the second inequality uses the fact that $\rank(\PlErrorA) \leq
2 \rdim$. \\

We now split into two cases, depending on whether or not the error
$\DeltaHat$ belongs to the set $\THMCONNEW$.

\paragraph{Case 1:} 
First suppose that $\DeltaHat \notin \THMCONNEW$.  In this case, by
the definition~\eqref{EqnDefnConstraintNew}, we have
\begin{align*}
\matsnorm{\DeltaHat}{F}^2 & \leq \Kcon \, \big(\MTIMES
\|\DeltaHat\|_\infty \big) \matsnorm{\DeltaHat}{1} \,
\sqrt{\frac{\mdim \log \mdim}{\numobs}} \\
& \leq 2 \Kcon \upperspike \matsnorm{\DeltaHat}{1} \,
\sqrt{\frac{\mdim \log \mdim}{\numobs}},
\end{align*}
since $\|\DeltaHat\|_\infty \leq \|\GammaStar\|_\infty +
\|\GammaHat\|_\infty \leq \frac{2 \upperspike}{\MTIMES}$.  Now
applying the bound~\eqref{EqnRestrictedTwo}, we obtain
\begin{align}
\label{EqnUpOne}
\matsnorm{\DeltaHat}{F}^2 & \leq 2 \Kcon \, \upperspike \,
\sqrt{\frac{\mdim \log \mdim}{\numobs}} \big \{ 8 \sqrt{\rdim}
\matsnorm{\DeltaHat}{F} + 4 \TAILSUM \big \}.
\end{align}

\paragraph{Case 2:}
Otherwise, we must have $\DeltaHat \in \THMCONNEW$.  Recall the
reformulated lower bound~\eqref{EqnRSCNew}.  On one hand, if
$\frac{128 \altwospike(\DeltaHat)}{\sqrt{\numobs}} > 1/2$, then we
have
\begin{align}
\label{EqnUpTwo}
\matsnorm{\DeltaHat}{F} & \leq \frac{256 \MTIMES
\|\DeltaHat\|_\infty}{\sqrt{\numobs}} \; \leq \; \frac{512
\spikeupper}{\sqrt{\numobs}}.
\end{align}
On the other hand, if $\frac{128
  \altwospike(\DeltaHat)}{\sqrt{\numobs}} \leq 1/2$, then
from the bound~\eqref{EqnRSCNew}, we have
\begin{align}
\label{EqnUseRSC}
\frac{\|\XopNew(\DeltaHat)\|_2}{\sqrt{\numobs}} & \geq
\frac{\matsnorm{\DeltaHat}{F}}{16}
\end{align}
with high probability.  Note that $\GammaHat$ is optimal and
$\GammaStar$ is feasible for the convex program~\eqref{EqnGammaSDP},
so that we have the basic inequality
\begin{align*}
\frac{1}{2 \numobs} \|y - \XopNew(\GammaHat)\|_2^2 + \regpar
\matsnorm{\GammaHat}{1} & \leq \frac{1}{2 \numobs} \|y -
\XopNew(\GammaStar)\|_2^2 + \regpar \matsnorm{\GammaStar}{1}.
\end{align*}
Some algebra yields
\begin{align*}
\frac{1}{2 \numobs} \|\XopNew(\DeltaHat)\|_2^2 & \leq \noisevar \,
\tracer{\DeltaHat}{\frac{1}{\numobs} \sum_{i=1}^\numobs \newvsca_i
\ObsTil{i}} + \regpar \matsnorm{\GammaStar + \DeltaHat}{1} - \regpar
\matsnorm{\DeltaHat},
\end{align*}
Substituting the lower bound~\eqref{EqnUseRSC} into this inequality
yields
\begin{align*}
\frac{\|\DeltaHat\|_F^2}{512} & \leq \noisevar \,
\tracer{\DeltaHat}{\frac{1}{\numobs} \sum_{i=1}^\numobs \newvsca_i
  \ObsTil{i}} + \regpar \matsnorm{\GammaStar + \DeltaHat}{1} - \regpar
\matsnorm{\DeltaHat}.
\end{align*}
From this point onwards, the proof is identical (apart from constants)
to Theorem 1 in Negahban and Wainwright~\cite{NegWai09}, and we obtain
that there is a numerical constant $\plaincon_1$ such that
\begin{align}
\label{EqnUpThree}
\matsnorm{\PlError}{F}^2 & \leq \plaincon_1 \, \spikeupper \, \regpar \biggr \{
\sqrt{\rdim} \matsnorm{\PlError}{F} + \TAILSUM \biggr\}.
\end{align}

\paragraph{Putting together the pieces:}  Summarizing our results,
we have shown that with high probability, one of the three
bounds~\eqref{EqnUpOne}, ~\eqref{EqnUpTwo} or~\eqref{EqnUpThree} must
hold.  Since $\upperspike \geq 1$, we can summarize by claiming that
there is a universal constant $\plaincon_1$ such that
\begin{align*}
\matsnorm{\PlError}{F}^2 & \leq \plaincon_1 \; \max \biggr \{ \regpar,
\sqrt{\frac{\mdim \log \mdim}{\numobs}} \biggr \} \; \big [
\sqrt{\rdim} \matsnorm{\PlError}{F} + \TAILSUM \big].
\end{align*}
Translating this result back to the original co-ordinate system
($\GammaStar = \sqrt{\Row} \ThetaStar \sqrt{\Col}$) yields the
claim~\eqref{EqnMatrixEst}.

\subsection{Proof of Corollary~\ref{CorExact}}

When $\ThetaStar$ (and hence $\sqrt{\Row} \ThetaStar \sqrt{\Col}$) has
rank $\rdim < \mrow$, then we have $\sum_{j = \rdim + 1}^\mrow
\sigma_j(\sqrt{\Row} \ThetaStar \sqrt{\Col}) = 0$.  Consequently, the
bound~\eqref{EqnMatrixEst} reduces to $\Weifro{\DeltaTil} \leq
\plaincon_1 \, \upperspike \, \lamstar \sqrt{\rdim}$.
To complete the proof, it suffices to show that
\begin{align*}
\mprob \big[ \matsnorm{\frac{1}{\numobs} \sum_{i=1}^\numobs \newvsca_i
    \Row^{-1/2} \ObsMat{i} \Col^{-1/2}}{2} \geq \plaincon_1 \,
  \noisevar \, \sqrt{\frac{\mdim \log \mdim}{\numobs}} \big] & \leq
\plaincon_2 \exp(-\plaincon_2 \mdim \log \mdim).
\end{align*}
We do so via the Alhswede-Winter matrix bound, as stated in
Appendix~\ref{AppLemAW}.  Defining the random matrix \mbox{$\ObsY{i}
  \defn \newvsca_i \Row^{-1/2} \, \ObsMat{i} \, \Col^{-1/2}$}, we
first note that $\newvsca_i$ is sub-exponential with parameter $1$,
and $|\Row^{-1/2} \ObsMat{i} \Col^{-1/2}|$ has a single entry with
magnitude at most $\Lcon \MTIMES$, which implies that
\begin{align*}
\|\ObsY{i}\|_{\psi_1} \leq \Lcon \, \noisevar \, \MTIMES \leq 2
\noisevar \, \Lcon \mdim
\end{align*}
(Here $\|\cdot\|_{\psi_1}$ denotes the Orlicz norm~\cite{LedTal91} of
a random variable, as defined by the function $\psi_1(x) = \exp(x)
-1$; see Appendix~\ref{AppLemAW}).  Moreover, we have
\begin{align*}
\Exs[(\ObsY{i})^T \ObsY{i}] & = \noisevar^2 \, 
\Exs \Big[ \frac{\mrow
    \mcol}{\Row_{j(i)} \, \Col_{k(i)}} e_{k(i)} e_{j(i)}^T e_{j(i)}
  e_{k(i)}^T \Big] \\
& = \noisevar^2 \, \Exs \Big[ \frac{\mrow
    \mcol}{\Row_{j(i)} \, \Col_{k(i)}} e_{k(i)}
  e_{k(i)}^T \Big]   \\
& = \noisevar^2 \, \mrow I_{\mcol \times \mcol}.
\end{align*}
so that $\matsnorm{\Exs[(\ObsY{i})^T \ObsY{i}]}{2} \leq 2 \noisevar^2
\mdim$, recalling that $2 \mdim = \mrow + \mcol \geq \mrow$.  The same
bound applies to $\matsnorm{\Exs[\ObsY{i}(\ObsY{i})^T]}{2}$, so that
applying Lemma~\ref{LemAW} with $t = \numobs \delta$, we conclude that
\begin{align*}
\mprob \big[ \matsnorm{ \frac{1}{\numobs} \sum_{i=1}^\numobs
    \newvsca_i \Row^{-1/2} \ObsMat{i} \Col^{-1/2}}{2} \geq \delta
  \big] & \leq (\mrow \times \mcol) \, \max \big \{ \exp(-\numobs
\delta^2/(16 \noisevar^2 \mdim ), \; \exp(- \frac{\numobs \delta}{4
  \noisevar \, \Lcon \mdim}) \big \}
\end{align*}
Since $\MTIMES \leq \mrow + \mcol = 2 \mdim$, if we set $\delta^2 =
\plaincon_1^2 \noisevar^2 \frac{\mdim \log \mdim}{\numobs}$ for a
sufficiently large constant $\plaincon_1$, the result follows.  (Here
we also use the assumption that $\numobs = \Omega(\mdim \log \mdim)$,
so that the term $\sqrt{\frac{\mdim \log \mdim}{\numobs}}$ is
dominant.)

\subsection{Proof of Corollary~\ref{CorApprox}}

For this corollary, we need to determine an appropriate choice of
$\rdim$ so as to optimize the bound~\eqref{EqnMatrixEst}.  To ease
notation, let us make use of the shorthand notation $\GammaStar =
\sqrt{\Row} \ThetaStar \sqrt{\Col}$.  With the singular values of
$\GammaStar$ ordered in non-increasing order, fix some threshold $\tau
> 0$ to be determined, and set $\rdim = \max \{ j \, \mid \,
\sigma_j(\GammaStar) > \tau \}$.  This choice ensures that
\begin{align*}
\sum_{j=\rdim+1}^\mrow \sigma_j(\GammaStar) & = \tau \;
 \sum_{j=\rdim+1}^\mrow \frac{\sigma_j(\GammaStar)}{\tau} \; \leq \;
 \tau \sum_{j=\rdim+1}^\mrow
 \big(\frac{\sigma_j(\GammaStar)}{\tau}\big)^\qpar \; \leq \;
 \tau^{1-\qpar} \radq.
\end{align*}
Moreover, we have $\rdim \, \tau^\qpar \leq \sum_{j=1}^\rdim \big \{
\sigma_j(\GammaStar) \big\}^\qpar \; \leq \radq$, which implies that
$\sqrt{\rdim} \leq \sqrt{\radq} \tau^{-\qpar/2}$.  Substituting these
relations into the upper bound~\eqref{EqnMatrixEst} leads to
\begin{align*}
\Weifrosq{\DeltaTil} & \leq \plaincon_1 \; \upperspike \; \regpar^* \;
\; \big [ \sqrt{\radq} \tau^{-\qpar/2} \Weifro{\DeltaTil} +
\tau^{1-\qpar} \radq \big \}
\end{align*}
In order to obtain the sharpest possible upper bound, we set $\tau =
\upperspike \regpar^*$.  Following some algebra, we find that there is
a universal constant $\plaincon_1$ such that
\begin{align*}
\Weifrosq{\DeltaTil} & \leq \plaincon_1 \radq \; \big( (\upperspike)^2
(\lamstar)^2 \big)^{1 - \frac{\qpar}{2}}.
\end{align*}
As in the proof of Corollary~\ref{CorExact}, it suffices to choose
$\regpar = \Omega (\noisevar \sqrt{\frac{\mdim \log \mdim}{\numobs}})$,
so that \mbox{$\regpar^* = \order \sqrt{(\noisevar^2+1) \frac{\mdim \log
\mdim}{\numobs}})$,} from which the claim follows.


\subsection{Proof of Theorem~\ref{ThmLower}}

Our proof of this lower bound based on a combination of
\mbox{information-theoretic methods~\cite{Yu97,YanBar99},} which allow
us to reduce to a multiway hypothesis test, and an application of the
probabilistic method so as to construct a suitably large packing set.
By Markov's inequality, it suffices to prove that
\begin{align*}
\sup_{\ThetaStar \in \Ballspec(\radq)} \: \mprob \biggr[
  \matsnorm{\ThetaHat - \ThetaStar}{F}^2 \geq 2 \delta \biggr] & \geq
  \frac{1}{2}.
\end{align*}
In order to do so, we proceed in a standard way---namely, by reducing
the estimation problem to a testing problem over a suitably
constructed packing set contained within $\Ballspec(\radq)$.  In
particular, consider a set $\{\Theta^1, \ldots,
\Theta^{\PackNum(\delta)} \}$ of matrices, contained within
$\Ballspec(\radq)$, such that $\matsnorm{\Theta^k - \Theta^\ell}{F}
\geq \delta$ for all $\ell \neq k$.  To ease notation, we use
$\PackNum$ as shorthand for $\PackNum(\delta)$ through much of the
argument.  Suppose that we choose an index $\Lind \in \{1, 2, \ldots,
\PackNum\}$ uniformly at random (u.a.r.), and we are given
observations $y \in \real^\numobs$ from the observation
model~\eqref{EqnNetflix} with $\ThetaStar = \Theta^\Lind$.  Then
triangle inequality yields the lower bound
\begin{align*}
\sup_{\ThetaStar \in \Ballspec(\radq)} \mprob \biggr[
  \matsnorm{\ThetaHat - \ThetaStar}{F} \geq \delta \biggr] & \geq
  \mprob[ \Lhat \neq \Lind].
\end{align*}

If we condition on $\Xop$, a variant of Fano's inequality yields
\begin{align}
\label{EqnInitFano}
\mprob[ \Lhat \neq \Lind \, \mid \, \Xop] & \geq 1 - \frac{ ({\PackNum
    \choose 2})^{-1} \sum_{\ell \neq k} \kull{\Theta^k}{\Theta^\ell} +
    \log 2}{ \log \PackNum},
\end{align}
where $\kull{\Theta^k}{\Theta^\ell}$ denotes the Kullback-Leibler
divergence between the distributions of $(y | \Xop, \Theta^k)$ and $(y
| \Xop, \Theta^\ell)$.  In particular, for additive Gaussian noise
with variance $\noisevar^2$, we have
\begin{align*}
\kull{\Theta^k}{\Theta^\ell} & = \frac{1}{2 \noisevar^2} \|
\Xop(\Theta^k) - \Xop(\Theta^\ell)\|_2^2,
\end{align*}
and moreover,
\begin{align*}
\Exs_{\Xop} \big[ \kull{\Theta^k}{\Theta^\ell} \big] & = \frac{1}{2
\noisevar^2} \matsnorm{\Theta^k - \Theta^\ell}{F}^2.
\end{align*}
Combined with the bound~\eqref{EqnInitFano}, we obtain the bound
\begin{align}
\mprob[\Lhat \neq \Lind] & = \Exs_\Xop \big \{\mprob[ \Lhat \neq \Lind
    \, \mid \, \Xop] \big \} \nonumber \\
\label{EqnFanoTwo}
& \geq 1 - \frac{ ({\PackNum \choose 2})^{-1} \sum_{\ell \neq k}
     \frac{\numobs}{2 \noisevar^2} \matsnorm{\Theta^k -
     \Theta^\ell}{F}^2 + \log 2}{ \log \PackNum},
\end{align}

The remainder of the proof hinges on the following technical lemma,
which we prove in Appendix~\ref{AppLemPacking}.
\blems
\label{LemPacking}
Let $\mdim \geq 10$ be a positive integer, and let $\delta > 0$.  Then
for each $\rdim = 1, 2, \ldots, \mdim$, there exists a set of
$\mdim$-dimensional matrices $\{\Theta^1, \ldots, \Theta^{M} \}$ with
cardinality \mbox{$M = \lfloor \frac{1}{4} \exp \big( \frac{\rdim
\mdim}{128} \big) \rfloor$} such that each matrix has rank $\rdim$,
and moreover
\begin{subequations}
\begin{align}
\label{EqnPackFrob}
\matsnorm{\Theta^\ell}{F} & = \delta \qquad \mbox{for all $\ell = 1,
  2, \ldots, M$,} \\
\label{EqnPackSpread}
\matsnorm{\Theta^\ell - \Theta^k}{F} & \geq \delta \qquad \mbox{for
  all $\ell \neq k$,} \\
\label{EqnPackSpike}
\alspike(\Theta^\ell) & \leq \sqrt{32 \log \mdim} \quad \mbox{for all
  $\ell = 1, 2, \ldots, M$, and} \\
\label{EqnPackRank}
\matsnorm{\Theta^\ell}{op} & \leq \frac{4 \delta}{\sqrt{r}} \quad
\mbox{for all $\ell = 1, 2, \ldots, M$.}
\end{align}
\end{subequations}
\elems

We now show how to use this packing set in our Fano bound.  To avoid
technical complications, we assume throughout that $\rdim \mdim > 1024
\log 2$.  Note that packing set from Lemma~\ref{LemPacking} satisfies
$\matsnorm{\Theta^k - \Theta^\ell}{F} \leq 2 \delta$ for all $k \neq
\ell$, and hence from Fano bound~\eqref{EqnFanoTwo}, we obtain
\begin{align*}
\mprob[\Lhat \neq \Lind] & \geq 1 - \frac{2 \frac{\numobs
  \delta^2}{\noisevar^2} + \log 2}{\frac{\rdim \mdim}{128} - \log 4}
  \\
& \geq 1 - \frac{2 \frac{\numobs \delta^2}{\noisevar^2} \ + \log
    2}{\frac{\rdim \mdim}{256}} \\
& \geq 1 - \frac{512 \frac{\numobs \delta^2}{\noisevar^2} + 256 \log
2}{\rdim \mdim}.
\end{align*}
If we now choose $\delta^2 = \frac{\noisevar^2}{2048} \frac{\rdim
  \mdim}{\numobs}$, then
\begin{align*}
\mprob[\Lhat \neq \Lind] & \geq 1 - \frac{ \frac{\rdim \mdim}{4} + 256
  \log 2}{\rdim \mdim} \; \geq \; \frac{1}{2},
\end{align*}
where the final inequality again uses the bound $\rdim \mdim \geq 1024
\log 2$.

In the special case $\qpar = 0$, the proof is complete, since the
elements $\Theta^\ell$ all have rank $r = R_0$, and satisfy the bound
$\alspike(\Theta^\ell) \leq \sqrt{32 \log \mdim}$.  For $\qpar \in
(0,1]$, consider the matrix class $\Ballspec(\radq)$, and let us set
$\rdim = \min \{ \mdim, \lceil \radq \big(\frac{\mdim}{\numobs}
\big)^{-\frac{\qpar}{2}} \rceil \}$ in Lemma~\ref{LemPacking}.  With
this choice, since each matrix $\Theta^\ell$ has rank $\rdim$, we have
\begin{align*}
\sum_{j=1}^\pdim \sigma_i(\Theta^\ell)^\qpar & \leq \; \rdim \,
\biggr(\frac{\delta}{\sqrt{\rdim}}\biggr)^{\qpar} \; = \; \rdim
\biggr(\frac{1}{2048} \sqrt{\frac{\mdim}{\numobs}} \biggr)^{\qpar} \; \leq
\; \radq,
\end{align*}
so that we are guaranteed that $\Theta^\ell \in \Ballspec(\radq)$.
Finally, we note that
\begin{align*}
\frac{\rdim \mdim}{\numobs} & \geq \min \big \{ \radq \, \biggr(
\frac{\mdim}{\numobs} \biggr)^{1-\frac{\qpar}{2}}, \;
\frac{\mdim^2}{\numobs} \big \},
\end{align*}
so that we conclude that the minimax error is lower bounded by
\begin{align*}
\frac{1}{4096} \min \biggr \{ \radq \, \biggr(\frac{\noisevar^2
\mdim}{\numobs} \biggr)^{1-\frac{\qpar}{2}}, \; \frac{\noisevar^2
\mdim^2}{\numobs} \biggr \}
\end{align*}
 for $\mdim \rdim$ sufficiently large.  (At the expense of a worse
 pre-factor, the same bound holds for all $\mdim \geq 10$.)
%


\section{Proof of Theorem~\ref{ThmRSC}}
\label{SecProofThmRSC}

We now turn to the proof that the sampling operator in weighted matrix
completion satisfies restricted strong convexity over the set
$\Constraint$, as stated in Theorem~\ref{ThmRSC}. In order to lighten
notation, we prove the theorem in the case $\mrow = \mcol$.  In terms
of rates, this is a worst-case assumption, effectively amounting to
replacing both $\mrow$ and $\mcol$ by the worst-case $\max \{ \mrow,
\mcol \}$.  However, since our rates are driven by $\mdim =
\frac{1}{2} (\mrow + \mcol)$ and we have the inequalities
\begin{equation*}
\frac{1}{2} \max \{ \mrow, \mcol \} \; \leq \frac{1}{2} (\mrow +
\mcol) \; \leq \max \{ \mrow, \mcol\},
\end{equation*}
this change has only an effect on the constant factors.  The proof can
be extended to the general setting $\mrow \neq \mcol$ by appropriate
modifications if these constant factors are of interest.


\subsection{Reduction to simpler events}

In order to prove Theorem~\ref{ThmRSC}, it is equivalent to show that,
with high probability, we have
\begin{align}
\label{EqnRapBound}
\frac{\|\XopNew(\GamPlain)\|_2}{\sqrt{\numobs}} & \geq \frac{1}{8}
\matsnorm{\GamPlain}{F} - \frac{\LeadConHalf \; \MTIMESPOST \;
\|\GamPlain\|_\infty}{ \sqrt{\numobs}} \qquad \mbox{for all $\GamPlain
\in \THMCONNEW$.}
\end{align}
The remainder of the proof is devoted to studying the ``bad'' event
\begin{align}
\EvilEvent(\XopNew) & \defn \biggr \{ \exists \quad \GamPlain \in
\THMCONNEW \, \mid \,
\Big|\frac{\|\XopNew(\GamPlain)\|_2}{\sqrt{\numobs}} -
\matsnorm{\GamPlain}{F} \Big| > \frac{7}{8} \matsnorm{\GamPlain}{F} +
\frac{\LeadConHalf \; \MTIMESPOST \; \|\GamPlain\|_\infty}{ \sqrt{\numobs}}
\biggr \}.
\end{align}
Suppose that $\EvilEvent(\XopNew)$ does \emph{not} hold: then we have
\begin{align*}
\Big|\frac{\|\XopNew(\GamPlain)\|_2}{\sqrt{\numobs}} -
\matsnorm{\GamPlain}{F} \Big| & \leq \frac{7}{8}
\matsnorm{\GamPlain}{F} + \frac{\LeadConHalf \; \MTIMESPOST \;
\|\GamPlain\|_\infty}{ \sqrt{\numobs}} \qquad \mbox{for all $\GamPlain
\in \THMCONNEW$,}
\end{align*}
which implies that the bound~\eqref{EqnRapBound} holds.  Consequently,
in terms of the ``bad'' event, the claim of Theorem~\ref{ThmRSC} is
implied by the tail bound $\mprob[\EvilEvent(\XopNew)] \leq 16
\exp(-\plaincon' \MSUM \log \MSUM )$.

We now show that in order to establish a tail bound on
$\EvilEvent(\XopNew)$, it suffices to bound the probability of some
simpler events $\Event(\XopNew; \myrad)$, defined below.  Since the
definition of the set $\THMCONNEW$ and event $\EvilEvent(\XopNew)$ is
invariant to rescaling of $\GamPlain$, we may assume without loss of
generality that $\|\GamPlain\|_\infty = \frac{1}{\MTIMESPOST}$.  The
remaining degrees of freedom in the set $\THMCONNEW$ can be
parameterized in terms of the quantities $\myrad =
\matsnorm{\GamPlain}{F}$ and $\SpecRadPlain =
\matsnorm{\GamPlain}{1}$.  For any $\GamPlain \in \THMCONNEW$ with
$\|\GamPlain\|_\infty = \frac{1}{\MTIMESPOST}$ and
$\matsnorm{\GamPlain}{F} \leq \myrad$, we have
$\matsnorm{\GamPlain}{1} \leq \SpecRad$, where
\begin{align}
\label{EqnDefnRho}
\SpecRad & \defn \frac{\myrad^2}{\Kcon \, \LowerMess}.
\end{align}

For each radius $\myrad > 0$, consider the set
\begin{align}
\label{EqnDefnSpecSet}
\SpecSet(\myrad) & \defn \big \{ \GamPlain \in \THMCONNEW \, \mid \,
\|\GamPlain\|_\infty = \frac{1}{\MTIMESPOST}, \; \matsnorm{\GamPlain}{F}
\leq \myrad, \; \matsnorm{\GamPlain}{1} \leq \SpecRad \big \},
\end{align}
and the associated event
\begin{align}
\Event(\XopNew; \myrad) & \defn \biggr \{ \exists \quad \GamPlain \in
\SpecSet(\myrad) \, \mid \,
\big|\frac{\|\XopNew(\GamPlain)\|_2}{\sqrt{\numobs}} -
\matsnorm{\GamPlain}{F} \big| \geq \frac{3}{4} \myrad + \frac{ \TEA
  L}{\sqrt{\numobs}} \biggr \}.
\end{align}
The following lemma shows that it suffices to upper bound the
probability of the event $\Event(\XopNew; \myrad)$ for each fixed
$\myrad > 0$.
\blems
\label{LemPeeling}
Suppose that are universal constants $(\plaincon_1, \plaincon_2)$ such
that
\begin{align}
\label{EqnTailBoundEvent}
\mprob[\Event(\XopNew; \myrad)] & \leq \plaincon_1 \exp(-\plaincon_2
\numobs \myrad^2)
\end{align}
for each fixed $\myrad > 0$.  Then there is a universal constant
$\plaincon'_2$ such that
\begin{align}
\mprob[\EvilEvent(\XopNew)] & \leq \plaincon_1
  \frac{\exp(-\plaincon'_2 \MSUM \log \MSUM)}{1 - \exp(-\plaincon'_2
  \MSUM \log \MSUM)}.
\end{align}
\elems
\noindent The proof of this claim, provided in
Appendix~\ref{AppLemPeeling}, follows by a peeling argument.

\subsection{Bounding the probability of $\Event(\XopNew; \myrad)$}

Based on Lemma~\ref{LemPeeling}, it suffices to prove the tail
bound~\eqref{EqnTailBoundEvent} on the event $\Event(\XopNew; \myrad)$
for each fixed $\myrad > 0$.  Let us define
\begin{align}
 Z_\numobs(\myrad) & \defn \sup_{\GamPlain \in \SpecSetTil(\myrad)}
 \Biggr | \SuperSum{\GamPlain} - \matsnorm{\GamPlain}{F} \Biggr|,
\end{align}
where 
\begin{align}
\label{EqnDefnSpecsetTil}
\SpecSetTil(\myrad) & \defn \big \{ \GamPlain \in \THMCONNEW \, \mid
\, \|\GamPlain\|_\infty \leq \frac{1}{\MTIMESPOST}, \;
\matsnorm{\GamPlain}{F} \leq \myrad, \; \matsnorm{\GamPlain}{1} \leq
\SpecRad \big \}.
\end{align}
(The only difference from $\SpecSet(\myrad)$ is that we have relaxed
to the inequality $\|\GamPlain\|_\infty \leq \frac{1}{\MTIMESPOST}$.)
In the remainder of this section, we prove that there are universal
constants $(\plaincon_1, \plaincon_2)$ such that
\begin{align}
\label{EqnZtail}
\mprob \big[Z_\numobs(\myrad) \geq \frac{3}{4} \myrad +
  \frac{\LeadConHalf}{\sqrt{\numobs}} \big] & \leq \plaincon_1 \exp(-
\plaincon_2 \frac{\numobs \myrad^2}{\Lcon^2}) \quad \mbox{for each
  fixed $\myrad > 0$.}
\end{align}
This tail bound means that the condition of Lemma~\ref{LemPeeling} is
satisfied, and so completes the proof of Theorem~\ref{ThmRSC}.

\vspace*{.2in}

In order to prove~\eqref{EqnZtail}, we begin with a discretization
argument.  Let $\GamPlain^1, \ldots, \GamPlain^{N(\delta)}$ be a
$\delta$-covering of $\SpecSetTil(\myrad)$ in the Frobenius norm.  By
definition, given an arbitrary $\GamPlain \in \SpecSetTil(\myrad)$,
there exists some index $k \in \{1, \ldots, N(\delta) \}$ and a matrix
$\Delta \in \real^{\prow \times \pcol}$ with $\matsnorm{\Delta}{F}
\leq \delta$ such that $\GamPlain = \GamPlain^k + \Delta$. Therefore,
we have
\begin{eqnarray*}
\SuperSum{\GamPlain} - \matsnorm{\GamPlain}{F} & = &
\SuperSum{\GamPlain^k + \Delta} - \matsnorm{\GamPlain^k + \Delta}{F}
\\
& \leq & \SuperSum{\GamPlain^k} + \SuperSum{\Delta} -
\matsnorm{\GamPlain^k}{F} + \matsnorm{\Delta}{F} \\
& \leq & \Big|\SuperSum{\GamPlain^k} - \matsnorm{\GamPlain^k}{F} \Big| +
\SuperSum{\Delta} + \delta,
\end{eqnarray*}
where we have used the triangle inequality.  Following the same steps
establishes that this inequality holds for the absolute value of the
difference.

Moreover, since $\Delta = \GamPlain^k - \GamPlain$ with both
$\GamPlain^k$ and $\GamPlain$ belonging to $\SpecSetTil(\myrad)$, we
have $\matsnorm{\Delta}{1} \leq 2 \SpecRadPlain(\myrad)$ and
$\|\Delta\|_\infty \leq \frac{2}{\MTIMESPOST}$, where we have used the
definition~\eqref{EqnDefnSpecSet}.  Putting together the pieces, we
conclude that
\begin{multline}
\label{EqnThreeUpper}
Z_\numobs(\myrad) \leq \delta + \max_{k=1, \ldots, N(\delta)}
\Big|\SuperSum{\GamPlain^k} - \matsnorm{\GamPlain^k}{F} \Big| +
\AnnoySup{\Delta} \big| \SuperSum{\Delta}\big|,
\end{multline}
where 
\begin{align}
\label{EqnDefnAnnoySet}
\AnnoySet & \defn \big \{ \Delta \in \real^{\prow \times \pcol} \,
\mid \matsnorm{\Delta}{F} \leq \delta, \, \matsnorm{\Delta}{1} \leq 2
\SpecRadPlain(\myrad), \; \|\Delta\|_\infty \leq \frac{2}{\MTIMESPOST}
\big \}.
\end{align}

Note that the bound~\eqref{EqnThreeUpper} holds for any choice of
$\delta > 0$.  We establish the tail bound~\eqref{EqnZtail} with the
choice $\delta = \myrad/8$, and using the following two lemmas.  The
first lemma provides control of the maximum over the covering set:
\blems
\label{LemInnerControl}
As long $\MSUM \geq 10$, we have
\begin{align}
\label{EqnInnerControl}
\max_{k=1, \ldots, N(\myrad/8)} \Big|\SuperSum{\GamPlain^k} -
\matsnorm{\GamPlain^k}{F} \Big| & \leq \frac{\myrad}{8} + \frac{\TEA 
  L}{\sqrt{\numobs}}
\end{align}
with probability greater than $1- c \exp \big( - \, \frac{\numobs
  \myrad^2}{2048 \, \Lcon^2}\big)$.
\elems
\noindent See Appendix~\ref{AppLemInnerControl} for the proof of 
this claim. \\

  Our second lemma, proved in Appendix~\ref{AppLemRudel}, provides
control over the final term in the upper bound~\eqref{EqnThreeUpper}.
\blems
\label{LemRudel}
\begin{align*}
\AnnoySupTwo{\Delta}{\frac{\myrad}{8}} \big| \SuperSum{\Delta}\big| &
\leq \frac{\myrad}{2}
\end{align*}
with probability at least $1- 2 \exp \big( -\frac{\numobs
  \myrad^2}{8192 \Lcon^2} \big)$.
\elems

Combining these two lemmas with the upper bound~\eqref{EqnThreeUpper}
with $\delta = \myrad/8$, we obtain
\begin{align*}
Z_\numobs(\myrad) & \leq \frac{\myrad}{8} + \frac{\myrad}{8} +
\frac{\TEA L }{\sqrt{\numobs}} + \frac{\myrad}{2} \\
& \leq \frac{3\myrad}{4}  + \frac{\TEA L}{\sqrt{\numobs}}
\end{align*}
with probability at least $1- 4 \exp \big( -\frac{\numobs
  \myrad^2}{8192} \big)$, thereby establishing the tail
  bound~\eqref{EqnZtail} and completing the proof of
  Theorem~\ref{ThmRSC}.


\section{Discussion}
\label{SecDiscuss}

In this paper, we have established error bounds for the problem of
weighted matrix completion based on partial and noisy observations.
We proved both a general result, one which applies to any matrix, and
showed how it yields corollaries for both the cases of exactly
low-rank and approximately low-rank matrices.  A key technical result
is establishing that the matrix sampling operator satisfies a suitable
form of restricted strong convexity~\cite{NegRavWaiYu09} over a set of
matrices with controlled rank and spikiness.  Since more restrictive
properties such as RIP do not hold for matrix completion, this RSC
ingredient is essential to our analysis.  Our proof of the RSC
condition relied on a number of techniques from empirical process and
random matrix theory, including concentration of measure, contraction
inequalities and the Ahlswede-Winter bound.  Using
information-theoretic methods, we also proved that up to logarithmic
factors, our error bounds cannot be improved upon by any algorithm,
showing that our method is essentially minimax-optimal.

There are various open questions that remain to be studied.  Although
our analysis applies to both uniform and non-uniform sampling models,
it is limited to the case where each row (or column) is sampled with a
certain probability.  It would be interesting to consider extensions
to settings in which the sampling probability differed from entry to
entry, as investigated empirically by Salakhutdinov and
Srebro~\cite{SalSre10}.

\subsection*{Acknowledgments}

SN and MJW were partially supported by NSF grants DMS-0907632,
NSF-CDI-0941742, and Air Force Office of Scientific Research
AFOSR-09NL184.


\appendix

\section{Proof of Lemma~\ref{LemPacking}}
\label{AppLemPacking}

We proceed via the probabilistic method, in particular by showing that
a random procedure succeeds in generating such a set with probability
at least $1/2$.  Let $M' = \exp \big(\frac{\rdim \mdim}{128} \big)$,
and for each $\ell = 1, \ldots, M'$, we draw a random matrix
$\ThetaInt^\ell \in \real^{\mdim \times \mdim}$ according to the
following procedure:
\begin{enumerate}
\item[(a)] For rows $i = 1, \ldots, \rdim$ and for each column $j = 1,
  \ldots, \mdim$, choose each $\ThetaInt^{\ell}_{ij} \in \{-1, +1\}$
  uniformly at random, independently across $(i,j)$.
\item[(b)] For rows $i = \rdim+1, \ldots, \mdim$, set
  $\ThetaInt^{\ell}_{ij} = 0$.
\end{enumerate}
We then let $Q \in \real^{\mdim \times \mdim}$ be a random unitary
matrix, and define $\Theta^\ell = \frac{\delta}{\sqrt{\rdim \mdim}} \:
Q \ThetaInt^\ell$ for all \mbox{$\ell = 1, \ldots, M'$.}  The
remainder of the proof analyzes the random set $\{\Theta^1, \ldots,
\Theta^{M'} \}$, and shows that it contains a subset of size at least
$M = M'/4$ that has properties (a) through (d) with probability at
least $1/2$.

By construction, each matrix $\ThetaInt^\ell$ has rank at most
$\rdim$, and Frobenius norm $\matsnorm{\ThetaInt^{\ell}}{F} =
\sqrt{\rdim \mdim}$.  Since $Q$ is unitary, the rescaled matrices
$\Theta^\ell$ have Frobenius norm $\matsnorm{\Theta^\ell}{F} =
\delta$.  We now prove that
\begin{align*}
\matsnorm{\Theta^\ell - \Theta^k}{F} & \geq \delta \qquad \mbox{for
  all $\ell \neq k$}
\end{align*}
with probability at least $1/8$.  Again, since $Q$ is unitary, it
suffices to show that \mbox{$\matsnorm{\ThetaInt^\ell -
    \ThetaInt^k}{F} \geq \sqrt{\rdim \mdim}$} for any pair $\ell \neq
k$.  We have
\begin{align*}
\frac{1}{\rdim \mdim} \matsnorm{\ThetaInt^k - \ThetaInt^\ell}{F}^2 & =
\frac{1}{\rdim \mdim} \sum_{i=1}^\rdim \sum_{j=1}^\mdim
\big(\ThetaInt^\ell_{ij} - \ThetaInt^k_{ij} \big)^2.
\end{align*}
This is a sum of $\rdim \mdim$ i.i.d. variables, each bounded by $4$.
The mean of the sum is $2$, so that the Hoeffding bound implies that
\begin{align*}
\mprob \big[ \frac{1}{\rdim \mdim} \matsnorm{\ThetaInt^k -
    \ThetaInt^\ell}{F}^2 \leq 2 - t \big] & \leq 2 \exp \big(-\rdim
\mdim \, t^2/32 \big).
\end{align*}
Since there are less than $(M')^2$ pairs of matrices in total, setting
$t = 1$ yields
\begin{align*}
\mprob \big[ \min_{\ell, k = 1, \ldots, M'}
  \frac{\matsnorm{\ThetaInt^\ell - \ThetaInt^k}{F}^2}{\rdim \mdim}
  \geq 1 \big] & \geq 1-2 \exp \big( - \frac{\rdim \mdim}{32} + 2 \log
M' \big) \; \geq \frac{7}{8},
\end{align*}
where we have used the facts $\log M' = \frac{\rdim \mdim}{128}$ and
$\mdim \geq 10$.  Recalling the definition of $\Theta^\ell$, we
conclude that
\begin{align}
\label{EqnInterWonZero}
\mprob \big[ \min_{\ell, k = 1, \ldots, M'} \matsnorm{\Theta^\ell -
    \Theta^k}{F}^2 \geq \delta^2 \big] & \geq \frac{7}{8}.
\end{align}

We now establish bounds on $\alspike(\Theta^\ell)$ and
$\matsnorm{\Theta^\ell}{2}$.  We first prove that for any fixed index
$\ell \in \{1, 2, \ldots, M'\}$, our construction satisfies
\begin{align}
\label{EqnInterWon}
\mprob \Big[ \alspike(\Theta^\ell) \leq \sqrt{32 \log \mdim} \Big]
\geq \frac{3}{4}.
\end{align}
Indeed, for any pair of indices $(i,j)$, we have $|\Theta^{\ell}_{ij}|
= |\inprod{q_i}{v_j}|$, where $q_i \in \real^\mdim$ is drawn from the
uniform distribution over the $\mdim$-dimensional sphere, and
$\|v_j\|_2 = \sqrt{\rdim} \; \frac{\delta}{\sqrt{\rdim \mdim}} =
\frac{\matsnorm{\Theta^\ell}{F}}{\sqrt{\mdim}}$.  By Levy's theorem
for concentration on the sphere~\cite{Ledoux01}, we have
\begin{align*}
\mprob \big[ |\inprod{q_i}{v_j}| \geq t \big] & \leq 2 \exp
\big(-\frac{\mdim^2}{8 \, \matsnorm{\Theta^\ell}{F}^2} t^2 \big).
\end{align*}
Setting $t = s/\mdim$ and taking the union bound over all $\mdim^2$
indices, we obtain
\begin{align*}
\mprob \big[ \mdim \, \|\Theta^\ell\|_\infty \geq s \big] & \leq 2
\exp \biggr( -\frac{1}{8 \, \matsnorm{\Theta^\ell}{F}^2} s^2 + 2 \log
\mdim \biggr).
\end{align*}
This probability is less than $1/2$ for $s = \matsnorm{\Theta^\ell}{F}
\, \sqrt{32 \log \mdim}$ and $\mdim \geq 2$, which establishes the
intermediate claim~\eqref{EqnInterWon}.  

Finally, we turn to property (d).  For each fixed $\ell$, by
definition of $\Theta^\ell$ and the unitary nature of $Q$, we have
$\matsnorm{\Theta^\ell}{op} = \frac{\delta}{\sqrt{\rdim \mdim}}
\matsnorm{U}{\ell}$, where $U \in \{-1, +1\}^{\rdim \times \mdim}$ is
a random matrix with i.i.d. Rademacher (and hence sub-Gaussian)
entries.  Known results on sub-Gaussian matrices~\cite{DavSza01} yield
\begin{align*}
\mprob \Big[\frac{\delta}{\sqrt{\rdim \mdim}} \matsnorm{U}{2} \leq
  \frac{2 \delta}{\sqrt{\rdim \mdim}} \big(\sqrt{\rdim} + \sqrt{\mdim}
  \big) \Big] & \geq 1 - 2 \exp\big(-\frac{1}{4} (\sqrt{\rdim} +
\sqrt{\mdim})^2 \big) \; \geq \; \frac{3}{4}
\end{align*}
for $\mdim \geq 10$.  Since $\rdim \leq \mdim$, we conclude that
\begin{align}
\label{EqnInterWonTwo}
\mprob \Big[\matsnorm{\Theta^\ell}{2} \leq \frac{4
    \delta}{\sqrt{\rdim}} \Big] & \geq \frac{3}{4}.
\end{align}

By combining the bounds~\eqref{EqnInterWon} and~\eqref{EqnInterWonTwo},
we find that for each fixed $\ell = 1, \ldots, M'$, we have
\begin{align}
\label{EqnInterWonThree}
\mprob \biggr[ \matsnorm{\Theta^\ell}{2} \leq \frac{4
    \delta}{\sqrt{\rdim}}, \;
  \frac{\alspike(\Theta^\ell)}{\matsnorm{\Theta}{F}} \leq \sqrt{32
    \log \mdim} \biggr] & \geq \frac{1}{2}
\end{align}
Consider the event $\Event$ that there exists a subset $S \subset \{1,
\ldots, M'\}$ of cardinality $M = \frac{1}{4} M'$ such that
\begin{equation*}
\matsnorm{\Theta^\ell}{2} \leq 4 \sqrt{\frac{\mdim}{\numobs}}, \quad
\mbox{and} \quad \frac{\alspike(\Theta^\ell)}{\matsnorm{\Theta}{F}}
\leq \sqrt{32 \log \mdim} \qquad \mbox{for all $\ell \in S$.}
\end{equation*}
By the bound~\eqref{EqnInterWonThree}, we have
\begin{align*}
\mprob[\Event] & \geq \sum_{k=M}^{M'} {M' \choose k} (1/2)^k.
\end{align*}
Since we have chosen $M < M'/2$, we are guaranteed that
$\mprob[\Event] \geq 1/2$, thereby completing the proof.

\section{Proof of Lemma~\ref{LemPeeling}}
\label{AppLemPeeling}

We first observe that for any $\GamPlain \in \THMCONNEW$ with
$\|\GamPlain\|_\infty = \frac{1}{\MTIMESPOST}$, we have
\begin{align*}
\matsnorm{\GamPlain}{F}^2 & \geq \Kcon \, \matsnorm{\GamPlain}{1} \,
\sqrt{\frac{\MSUM \log \MSUM}{\numobs}} \; \geq \Kcon
\matsnorm{\GamPlain}{F} \, \sqrt{\frac{\MSUM \log \MSUM}{\numobs}},
\end{align*}
whence $\matsnorm{\GamPlain}{F} \geq \Kcon \, \sqrt{\frac{\MSUM \log
\MSUM}{\numobs}}$.  Accordingly, recalling the
definition~\eqref{EqnDefnSpecSet}, it suffices to restrict our
attention to sets $\SpecSet(\myrad)$ with $\myrad \geq \mu \defn
\Kcon \sqrt{\frac{\MSUM \log \MSUM}{\numobs}}$.
For $\ell = 1, 2, \ldots$ and $\alpha = \frac{7}{6}$, define the sets
\begin{align}
\Coke_\ell & \defn \big \{ \GamPlain \in \THMCONNEW \mid \,
 \|\GamPlain\|_\infty = \frac{1}{\mdim}, \quad \alpha^{\ell-1} \mu
 \leq \matsnorm{\GamPlain}{F} \leq \alpha^\ell \mu, \mbox{ and }
 \matsnorm{\GamPlain}{1} \leq \rho(\alpha^\ell \mu) \big \}.
\end{align}
From the definition~\eqref{EqnDefnSpecSet}, note that by construction,
 we have $\Coke_\ell \subset \; \SpecSet(\alpha^\ell \mu)$.

Now if the event $\EvilEvent(\XopNew)$ holds for some matrix
$\GamPlain$, then this matrix $\GamPlain$ must belong to some set
$\Coke_\ell$.  When $\GamPlain \in \Coke_\ell$, then we are guaranteed
the existence of a matrix $\GamPlain \in \SpecSet(\alpha^\ell \mu)$
such that
\begin{align*}
\big| \frac{\|\XopNew(\GamPlain)\|_2}{\sqrt{\numobs}} -
\matsnorm{\GamPlain}{F} \big| & > \frac{7}{8} \matsnorm{\GamPlain}{F}
+ \frac{\LeadConHalf}{\sqrt{\numobs}} \\
 & \geq \frac{7}{8} \alpha^{\ell-1} \mu + \frac{\LeadConHalf}{\sqrt{\numobs}}
\\
& = \frac{3}{4} \alpha^\ell \mu + \frac{\LeadConHalf}{\sqrt{\numobs}},
\end{align*}
where the final equality follows since $\alpha = 7/6$.  Thus, we have
shown that when the violating matrix $\GamPlain \in \Coke_\ell$, then
event $\Event(\XopNew; \alpha^\ell \mu)$ must hold.  Since any
violating matrix must fall into some set $\Coke_\ell$, the union bound
implies that
\begin{align*}
\mprob[\EvilEvent(\XopNew)] & \leq \sum_{\ell=1}^\infty
\mprob[\Event(\XopNew; \alpha^\ell \mu)] \\
& \leq \plaincon_1 \sum_{\ell=1}^\infty \exp \big( - \plaincon_2
\numobs \alpha^{2 \ell} \mu^2 \big) \\
& \leq \plaincon_1 \, \sum_{\ell=1}^\infty \exp \big( - 2 \plaincon_2
\log(\alpha) \, \ell \, \numobs \mu^2 \big) \\ 
& \leq 4 \frac{\exp(-\plaincon'_2 \numobs \mu^2)}{1 -
\exp(-\plaincon'_2 \numobs \mu^2)}
\end{align*}
Since $\numobs \mu^2 = \Omega(\MSUM \log \MSUM)$, the claim follows.

\section{Proof of Lemma~\ref{LemInnerControl}}
\label{AppLemInnerControl}

For a fixed matrix $\GamPlain$, define the function
$F_\GamPlain(\XopNew) = \frac{1}{\sqrt{\numobs}}
\|\XopNew(\GamPlain)\|_2$.  We prove the lemma in two parts: first, we
establish that for any fixed $\GamPlain$, the function $F_\GamPlain$
satisfies the tail bound
\begin{align}
\label{EqnFixedTail}
\mprob \big[ |F_\GamPlain(\XopNew) - \matsnorm{\GamPlain}{F} | \geq
  \delta + \frac{\TEA \Lcon}{\sqrt{\numobs}} \big] & \leq 4 \exp \big( -
  \frac{\numobs \delta^2}{4 \Lcon^2} \big).
\end{align}
We then show that there exists a $\delta$-covering of
$\SpecSetTil(\myrad)$ such that
\begin{align}
\label{EqnCoverUpper}
\log N(\delta) & \leq 36 \big(\SpecRadPlain(\myrad)/\delta\big)^2 \;
\; \MSUMPOST.
\end{align}

Combining the tail bound~\eqref{EqnFixedTail} with the union bound, we
obtain
\begin{align*}
\mprob \big[ \max_{k = 1, \ldots, N(\delta)} |F_\GamPlain(\XopNew) -
  \matsnorm{\GamPlain^k}{F} | \geq \delta + \frac{16
  \Lcon}{\sqrt{\numobs}} \big] & \leq 4 \exp \big( - \frac{\numobs
  \delta^2}{4 \Lcon^2} + \log N(\delta) \big) \\
& \leq 4 \exp \biggr \{ - \frac{\numobs \delta^2}{4 \Lcon^2} + 36
\big(\SpecRadPlain(\myrad)/\delta\big)^2 \; \MSUMPOST \biggr \} \\
\end{align*}
where the final inequality follows uses the
bound~\eqref{EqnCoverUpper}.  Since Lemma~\ref{LemInnerControl} is
based on the choice $\delta = \myrad/8$, it suffices to show that
\begin{align*}
\frac{\numobs \myrad^2}{512 \, \Lcon^2} & \geq 36 \,
  \big(\SpecRadPlain(\myrad)/(\myrad/8) \big)^2 \; \MSUMPOST \\
& \stackrel{(a)}{=} 36 \, \biggr( \frac{8 \, \myrad}{\Kcon}
  \sqrt{\frac{\numobs}{\MSUM \log \MSUM}} \biggr)^2 \; \MSUMPOST \\
& = \frac{2304 \, \myrad^2}{\Kcon^2} \; \frac{\numobs}{\log \MSUM}.
\end{align*}
Noting that the terms involving $\myrad^2$ and $\numobs$ both cancel
out, we see that for any fixed $\Kcon$, this inequality holds once
$\log \MSUM$ is sufficiently large.  By choosing $\Kcon$ sufficiently
large, we can ensure that it holds for all $\MSUM \geq 2$. \\

It remains to establish the two intermediate
claims~\eqref{EqnFixedTail} and~\eqref{EqnCoverUpper}.

\paragraph{Upper bounding the covering number~\eqref{EqnCoverUpper}:}

We start by proving the upper bound~\eqref{EqnCoverUpper} on the
covering number.  To begin, let $\Ntil(\delta)$ denote the
$\delta$-covering number (in Frobenius norm) of the nuclear norm ball
$\Ball_1(\SpecRadPlain(\myrad)) = \big \{ \Delta \in \real^{\prow
  \times \pcol} \, \mid \, \matsnorm{\Delta}{1} \leq
\SpecRadPlain(\myrad) \big \}$, and let $N(\delta)$ be the covering
number of the set $\SpecSetTil(\myrad)$.  We first claim that
$N(\delta) \leq \Ntil(\delta)$.  Let $\{\Gamma^1, \ldots,
\Gamma^{\Ntil(\delta)} \}$ be a $\delta$-cover of
$\Ball_1(\SpecRadPlain(\myrad))$, From
equation~\eqref{EqnDefnSpecsetTil}, note that the set
$\SpecSetTil(\myrad)$ is contained within
$\Ball_1(\SpecRadPlain(\myrad))$; in particular, it is obtained by
intersecting the latter set with the set
\begin{align*}
\TempSet & \defn \big \{ \Delta \in \real^{\mdim \times \mdim} \, \mid
\, \|\Delta\|_\infty \leq \frac{1}{\MTIMESPOST}, \;
\matsnorm{\Delta}{F} \leq \myrad \big \}.
\end{align*}
Letting $\Pi_\TempSet$ denote the projection operator under Frobenius
norm onto this set, we claim that $\{\Pi_\TempSet(\Gamma^j), j = 1,
\ldots, \Ntil(\delta) \}$ is a $\delta$-cover of
$\SpecSetTil(\myrad)$.  Indeed, since $\TempSet$ is non-empty, closed
and convex, the projection operator is
non-expansive~\cite{Bertsekas_nonlin}, and thus for any $\Gamma \in
\SpecSetTil(\myrad) \subset \TempSet$, we have
\begin{align*}
\matsnorm{\Pi_\TempSet(\Gamma^j) - \Gamma}{F} \; = \;
\matsnorm{\Pi_\TempSet(\Gamma^j) - \Pi_\TempSet(\Gamma)}{F} & \leq
\matsnorm{\Gamma^j - \Gamma}{F},
\end{align*}
which establishes the claim.

We now upper bound $\Ntil(\delta)$.  Let $G \in \real^{\prow \times
  \pcol}$ be a random matrix with i.i.d. $N(0,1)$ entries.  By Sudakov
  minoration (cf. Theorem 5.6 in Pisier~\cite{Pisier89}), we have
\begin{align*}
\sqrt{\log \Ntil(\delta)} \; & \; \leq \frac{3}{\delta} \; \Exs \big[
  \sup_{\matsnorm{\Delta}{1} \leq \SpecRadPlain(\myrad)}
  \tracer{G}{\Delta}\big] \\
& \leq \frac{3 \SpecRadPlain(\myrad)}{\delta} \; \Exs \big[
\matsnorm{G}{2} \big],
\end{align*}
where the second inequality follows from the duality between the
nuclear and operator norms.  From known results on the operator norms
Gaussian random matrices~\cite{DavSza01}, we have the upper bound
\mbox{$\Exs[\matsnorm{G}{2}] \leq 2 \sqrt{\mdim}$,} so that
\begin{align*}
\sqrt{\log \Ntil(\delta)} \; & \leq \; \frac{6
\SpecRadPlain(\myrad)}{\delta} \; \mdim,
\end{align*}
thereby establishing the bound~\eqref{EqnCoverUpper}.

\paragraph{Establishing the tail bound~\eqref{EqnFixedTail}:}
Recalling the definition of the operator
$\XopNew$, we have
\begin{align*}
F_\GamPlain(\XopNew) & = \frac{1}{\sqrt{\numobs}} \big \{
\sum_{i=1}^\numobs \tracer{\ObsTil{i}}{\GamPlain}^2 \big \}^{1/2} \\
& = \frac{1}{\sqrt{\numobs}} \sup_{\|u\|_2 = 1} \sum_{i=1}^n u_i
\tracer{\ObsTil{i}}{\GamPlain} \\
& = \frac{1}{\sqrt{\numobs}} \sup_{\|u\|_2 = 1} \sum_{i=1}^n u_i Y_i
\end{align*}
where we have defined the random variables $Y_i \defn
\tracer{\ObsTil{i}}{\GamPlain}$.  Note that each $Y_i$ is zero-mean,
and bounded by $2 \Lcon$ since
\begin{align*}
|Y_i| & = |\tracer{\ObsTil{i}}{\GamPlain}| \\
& \leq \big( \sum_{a,b} |\ObsTil{i}|_{ab} \big) \; \|\GamPlain\|_\infty
\; \leq \; 2 \Lcon.
\end{align*}
where we have used the facts that $\|\GamPlain\|_\infty \leq
2/\MTIMESPOST$, and $\sum_{a,b} |\ObsTil{i}|_{ab} \leq \Lcon \; \MTIMESPOST$,
by definition of the matrices $\ObsTil{i}$.

  Therefore, applying Corollary 4.8 from Ledoux~\cite{Ledoux01}, we
conclude that
\begin{align*}
\mprob \big[ |F_\GamPlain(\XopNew) - \Exs[F_\GamPlain(\XopNew)] | \geq
  \delta + \frac{32 L}{\sqrt{\numobs}} \big] & \leq 4 \exp \big( -
\frac{\numobs \delta^2}{4 \Lcon^2} \big).
\end{align*}
The same corollary implies that
\begin{align*}
\big| \sqrt{\Exs[F^2_\GamPlain(\XopNew)]} - \Exs[F_\GamPlain(\XopNew)]
\big| & \leq \frac{16 \Lcon}{\sqrt{\numobs}}.
\end{align*}
Since $\Exs[F^2_\GamPlain(\XopNew)] = \matsnorm{\GamPlain}{F}^2$, the
tail bound~\eqref{EqnFixedTail} follows.

\section{Proof of Lemma~\ref{LemRudel}}
\label{AppLemRudel}

From the proof of Lemma~\ref{LemInnerControl}, recall the definition
$F_\GamPlain(\XopNew) = \frac{1}{\sqrt{\numobs}}
\|\XopNew(\GamPlain)\|_2$ where $\XopNew$ is the random sampling
operator defined by the $\numobs$ matrices $(\ObsTil{1}, \ldots,
\ObsTil{\numobs})$.  Using this notation, our goal is to bound the
function
\begin{align*}
G(\XopNew) & \defn \AnnoySup{\Delta} F_\Delta(\XopNew),
\end{align*}
where we recall that $\AnnoySet \defn \big \{ \Delta \in \real^{\mrow
  \times \mcol} \, \mid \matsnorm{\Delta}{F} \leq \delta, \,
  \matsnorm{\Delta}{1} \leq 2 \SpecRad, \; \|\Delta\|_\infty \leq
  \frac{2}{\MTIMESPOST} \big \}$.  \mbox{Ultimately,} we will set $\delta =
  \frac{\myrad}{8}$, but we use $\delta$ until the end of the proof
  for compactness in notation.

Our approach is a standard one: first show concentration of $G$ around
its expectation $\Exs[G(\XopNew)]$, and then upper bound the
expectation.  We show concentration via a bounded difference
inequality; since $G$ is a symmetric function of its arguments, it
suffices to establish the bounded difference property with respect to
the first co-ordinate.  In order to do so, consider a second operator
$\widetilde{\XopNew}$ defined by the matrices $(Z^{(1)}, \ObsTil{2},
\ldots, \ObsTil{\numobs})$, differing from $\XopNew$ only in the first
matrix.  Given the pair $(\XopNew, \widetilde{\XopNew})$, we have
\begin{align*}
G(\XopNew) - G(\wtil{\XopNew}) & = \AnnoySup{\Delta} F_\Delta(\XopNew)
- \AnnoySup{\Theta} F_\Theta(\wtil{\XopNew}) \\
& \leq \AnnoySup{\Delta} \big [F_\Delta(\XopNew) - F_\Delta(\wtil{\XopNew})
  \big] \\
& \leq \AnnoySup{\Delta} \frac{1}{\sqrt{\numobs}} \|\XopNew(\Delta) -
\wtil{\XopNew}(\Delta)\|_2 \\
& = \AnnoySup{\Delta} \frac{1}{\sqrt{\numobs}} 
\big|
\tracer{\ObsTil{1} - Z^{(1)}}{\Delta} \big|.
\end{align*}
For any fixed $\Delta \in \AnnoySet$, we have 
\begin{align*}
\big| \tracer{\ObsTil{1} - Z^{(1)}}{\Delta} \big| & \leq 2 \Lcon
\MTIMESPOST \, \|\Delta\|_\infty \; \leq \; 4 \Lcon,
\end{align*}
where we have used the fact that the matrix $\ObsTil{1} - Z^{(1)}$ is
non-zero in at most two entries with values upper bounded by $2 \Lcon
\MTIMESPOST$.  Combining the pieces yields $G(\XopNew) - G(\wtil{\XopNew})
\leq \frac{4 \Lcon}{\sqrt{\numobs}}$.  Since the same argument can be
applied with the roles of $\XopNew$ and $\wtil{\XopNew}$ interchanged,
we conclude that $|G(\XopNew) - G(\wtil{\XopNew})| \leq \frac{4
\Lcon}{\sqrt{\numobs}}$.  Therefore, by the bounded differences
variant of the Azuma-Hoeffding inequality~\cite{Ledoux01}, we have
\begin{align}
\label{EqnGconc}
\mprob \big[ |G(\XopNew) - \Exs[G(\XopNew)]| \geq t\big] & \leq 2 \exp
\big( -\frac{\numobs t^2}{32 \Lcon^2} \big).
\end{align}

Next we bound the expectation.  First applying Jensen's inequality, we
have
\begin{align*}
(\Exs[G(\XopNew)])^2 & \leq \Exs[G^2(\XopNew)] \\
& = \Exs \big[ \AnnoySup{\Delta} \SuperSumSq{\Delta} \big] \\
& = \Exs \biggr[ \AnnoySup{\Delta} \biggr \{ \frac{1}{\numobs}
  \sum_{i=1}^\numobs \big[ \tracer{\ObsTil{i}}{\Delta}^2- \Exs[
  \tracer{\ObsTil{i}}{\Delta}^2] \big] + \matsnorm{\Delta}{F}^2 \biggr
  \} \biggr] \\
& \leq \Exs \biggr[ \AnnoySup{\Delta} \biggr \{ \frac{1}{\numobs}
  \sum_{i=1}^\numobs \big[ \tracer{\ObsTil{i}}{\Delta}^2- \Exs[
  \tracer{\ObsTil{i}}{\Delta}^2] \big] \biggr \} \biggr] + \delta^2,
\end{align*}
where we have used the fact that $\Exs[ \tracer{\ObsTil{i}}{\Delta}^2
  = \matsnorm{\Delta}{F}^2 \leq \delta^2$.  Now a standard
  symmetrization argument~\cite{LedTal91} yields
\begin{align*}
\Exs_{\XopNew}[G^2(\XopNew)] \; \leq 2 \, \Exs_{\XopNew, \rade}
\big[\AnnoySup{\Delta} \frac{1}{\numobs} \sum_{i=1}^\numobs \rade_i
\tracer{\ObsTil{i}}{\Delta}^2 \big] + \delta^2,
\end{align*}
where $\{\rade_i\}_{i=1}^\numobs$ is an i.i.d. Rademacher sequence.
Since $|\tracer{\ObsTil{i}}{\Delta}| \leq 2 \Lcon$ for all $i$, the
Ledoux-Talagrand contraction inequality (p. 112, Ledoux and
Talagrand~\cite{LedTal91}) implies that
\begin{align*}
\Exs[G^2(\XopNew)] \; \leq 16 \Lcon \, \: \Exs \big[\AnnoySup{\Delta}
  \; \big \{\frac{1}{\numobs} \sum_{i=1}^\numobs \rade_i
  \tracer{\ObsTil{i}}{\Delta} \big \} \big] + \delta^2.
\end{align*}
By the duality between operator and nuclear norms, we have
\begin{align*}
\big|\frac{1}{\numobs} \sum_{i=1}^\numobs \rade_i
\tracer{\ObsTil{i}}{\Delta} \big| & \leq \matsnorm{\frac{1}{\numobs}
\sum_{i=1}^\numobs \rade_i \ObsTil{i}}{2} \; \matsnorm{\Delta}{1},
\end{align*}
and hence, since $\matsnorm{\Delta}{1} \leq \rho(\myrad)$ for all
$\Delta \in \AnnoySet$, we have
\begin{align}
\label{EqnAlmost}
\Exs[G^2(\XopNew)] & \leq \; 16 \, \Lcon \, \rho(\myrad) \; \Exs
\big[\matsnorm{\frac{1}{\numobs} \sum_{i=1}^\numobs \rade_i
\ObsTil{i}}{2} \big] + \delta^2.
\end{align}

\vspace*{.2in}

It remains to bound the operator norm $\Exs
\big[\matsnorm{\frac{1}{\numobs} \sum_{i=1}^\numobs \rade_i
\ObsTil{i}}{2} \big]$.  The following lemma, proved in
Appendix~\ref{AppLemExp}, provides a suitable upper bound:
\blems
\label{LemExp}
We have the upper bound
\begin{align}
\Exs \big[ \matsnorm{\frac{1}{\numobs} \sum_{i=1}^\numobs \rade_i
    \ObsTil{i}}{2} \big] & \leq 10 \, \max \big \{ \sqrt{ \frac{\Lcon
    \, \MSUM \log \MSUM}{\numobs}}, \; \frac{\Lcon \, \MSUM \log
    \MSUM}{\numobs} \big \}.
\end{align}
\elems
\noindent Thus, as long as $\numobs = \Omega(\MSUM \log \MSUM)$,
combined with the earlier bound~\eqref{EqnAlmost}, we conclude that
 \begin{align*}
 \Exs[G(\XopNew)] \leq \sqrt{\Exs[G^2(\XopNew)] } & \leq \big[160 \,
   \Lcon^2 \, \SpecRad \; \sqrt{ \frac{\MSUM \log \MSUM}{\numobs}} +
   \delta^2 \big]^{1/2},
 \end{align*}
using the fact that $\Lcon \geq 1$.  By definition of $\SpecRad$, we
have
\begin{align*}
160 \, \Lcon^2 \, \SpecRad \; \sqrt{ \frac{\MSUM \log \MSUM}{\numobs}}
& = \frac{160 \, \Lcon^2}{\Kcon} \myrad^2 \; \leq \big(\frac{5
  \myrad}{16} \big)^2,
\end{align*}
where the final inequality can be guaranteed by choosing $\Kcon$
sufficiently large.

Consequently, recalling our choice $\delta = \myrad/8$ and using the
inequality $\sqrt{a^2 + b^2} \leq |a| + |b|$, we obtain
\begin{align*}
\Exs[G(\XopNew)] \leq \frac{5}{16} \myrad + \frac{\myrad}{8} =
\frac{7}{16} \myrad.
\end{align*}
Finally, setting $t = \frac{\myrad}{16}$ in the concentration
bound~\eqref{EqnGconc} yields
\begin{align*}
G(\XopNew) & \leq \frac{\myrad}{16} + \frac{7}{16} \myrad =
\frac{\myrad}{2}
\end{align*}
with probability at least $1- 2 \exp \big( -c' \, \frac{\numobs
  \myrad^2}{\Lcon^2} \big)$ as claimed.

\section{Proof of Lemma~\ref{LemExp}}
\label{AppLemExp}

We prove this lemma by applying a form of Ahlwehde-Winter matrix
bound~\cite{AhlWin02}, as stated in Appendix~\ref{AppLemAW}, to the
matrix $\ObsY{i} \defn \rade_i \ObsTil{i}$.  We first compute the
quantities involved in Lemma~\ref{LemAW}.  Note that $\ObsY{i}$ is a
zero-mean random matrix, and satisfies the bound
\begin{align*}
\matsnorm{\ObsY{i}}{2} & = \MTIMESPOST \frac{1}{\sqrt{\Row_{j(i)}} \;
\sqrt{\Col_{k(i)}}} \matsnorm{ \rade_i \; e_{j(i)} \, e_{k(i)}^T}{2}
\; \leq \; \Lcon \, \MTIMESPOST.
\end{align*}
Let us now compute the quantities $\sigma_i$ in Lemma~\ref{LemAW}.  We
have
\begin{align*}
\Exs \big[(\ObsY{i}^T) \ObsY{i} \big] & = \Exs \biggr[
\frac{\MTIMESPOST^2}{\Row_{j(i)} \, \Col_{k(i)}} e_{k(i)} e_{k(i)}^T
\biggr] \; = \; \mdim I_{\mdim \times \mdim}
\end{align*}
and similarly, $\Exs \big[\ObsY{i} \, (\ObsY{i})^T \big] = \mdim
I_{\mdim \times \mdim}$, so that
\begin{align*}
\sigma_i^2 = \max \biggr \{ \matsnorm{\Exs \big[\ObsY{i} \,
(\ObsY{i})^T \big]}{2}, \; \matsnorm{\Exs \big[(\ObsY{i})^T \,
\ObsY{i} \big]}{2} \biggr \} \; = \; \mdim.
\end{align*}

Thus, applying Lemma~\ref{LemAW} yields the tail bound
\begin{align*}
\mprob \big[ \matsnorm{\sum_{i=1}^\numobs \rade_i \ObsTil{i}}{2} \geq
t \big] & \leq 2 \, \mdim \, \max \big \{ \exp(- \frac{t^2}{4 \numobs
\mdim}), \; \exp( - \frac{t}{2 \Lcon \mdim}) \big \}.
\end{align*}
Setting $t = \numobs \delta$, we obtain
\begin{align*}
\mprob \big[ \matsnorm{\frac{1}{\numobs} \sum_{i=1}^\numobs \rade_i
    \ObsTil{i}}{2} \geq 2 \Lcon \, \delta \big] & \leq 2 \mdim \, \max
    \big \{ \exp(- \frac{\numobs \delta^2}{4 \mdim}), \exp(-
    \frac{\numobs \delta}{2 \Lcon \mdim}) \big \}.
\end{align*}

Recall that for any non-negative random variable $T$, we have $\Exs[T]
= \int_0^\infty \mprob[T \geq s] ds$. Applying this fact to \mbox{$T
\defn \matsnorm{\frac{1}{\numobs} \sum_{i=1}^\numobs \rade_i
\ObsTil{i}}{2}$} and integrating the tail bound, we obtain
\begin{align*}
\Exs \big[ \matsnorm{\frac{1}{\numobs} \sum_{i=1}^\numobs \rade_i
    \ObsTil{i}}{2} \big] & \leq 10 \, \max \big \{ \sqrt{ \frac{ \mdim
    \log \mdim}{\numobs}}, \; \frac{\Lcon \mdim \log \mdim}{\numobs}
    \big \}, \\
& \leq 10 \, \max \big \{ \sqrt{ \frac{\Lcon \, \mdim \log
    \mdim}{\numobs}}, \; \frac{\Lcon \mdim \log \mdim}{\numobs} \big
    \},
\end{align*}
where the second inequality follows since $\Lcon \geq 1$.


\section{Ahlswede-Winter matrix bound}
\label{AppLemAW}

Here we state a Bernstein version of the Ahlswede-Winter tail
bound~\cite{AhlWin02} for the operator norm of a sum of random
matrices. The version here is a slight weakening (but sufficient for
our purposes) of a result due to Recht~\cite{Recht09}; we also refer
the reader to the notes of Vershynin~\cite{Ver09}, and the
strengthened results provided by Tropp~\cite{Tro10}.

 Let $\ObsY{i}$ be independent $\mrow \times \mcol$ zero-mean random
matrices such that $\matsnorm{\ObsY{i}}{2} \leq M$, and define
\begin{align*}
\sigma_i^2 & \defn \max \big \{ \matsnorm{\Exs[(\ObsY{i})^T
\ObsY{i}]}{2}, \quad \matsnorm{\Exs[\ObsY{i} (\ObsY{i})^T]}{2} \},
\end{align*}
as well as $\sigma^2 \defn \sum_{i=1}^\numobs \sigma^2_i$.
\blems
\label{LemAW}
We have
\begin{align}
\label{EqnAW}
\mprob \big[ \matsnorm{ \sum_{i=1}^\numobs \ObsY{i}}{2} \geq t \big] &
\leq (\mrow \times \mcol) \, \max \big \{ \exp(-t^2/(4 \sigma^2), \;
\exp(- \frac{t}{2 M}) \big \}
\end{align}
\elems
\noindent As noted by Vershynin~\cite{Ver09}, the same bound also
holds under the assumption that each $\ObsY{i}$ is sub-exponential
with parameter $M = \|\ObsY{i}\|_{\psi_1}$.  Here we are using the
Orlicz norm 
\begin{align*}
\|Z\|_{\psi_1} & \defn \inf \{ t > 0 \, \mid
\Exs[\psi(|Z|/t)] < \infty \},
\end{align*}
defined by the function $\psi_1(x) = \exp(x) -1$, as is appropriate
for sub-exponential variables (e.g., see the book~\cite{LedTal91}).



\end{document}